\documentclass[]{article}
\usepackage{enumitem}
\usepackage{multirow}
\usepackage{graphicx} 
\usepackage{mdsymbol}
\usepackage[a4paper, margin=1in]{geometry}
\usepackage{authblk}
\usepackage{xurl}
\usepackage[table]{xcolor}
\usepackage[most]{tcolorbox}
\usepackage{subcaption} %
\usepackage{booktabs,caption}
\usepackage[flushleft]{threeparttable}
\usepackage{natbib}
\usepackage{setspace}
\usepackage{hyperref}
\usepackage{cleveref}

\date{}
\begin{document}
\hypersetup{pdftitle=Shaping Sustainability Education in Electrical and Computer Engineering}
\title{\textbf{Shaping Sustainability Education in Electrical and Computer Engineering}}

\author[1]{Esther Roorda \thanks{Email: estherr@ece.ubc.ca}}

\author[1]{Sathish Gopalakrishnan\thanks{Email: sathish@ece.ubc.ca}}
\date{November 23, 2026}

\affil[1]{The University of British Columbia, Vancouver, Canada}

\maketitle
\textbf{Keywords}: Electrical Engineering | Computer Engineering | Faculty attitudes | Institutional change | Mixed methods research | Sustainability 

\newcommand\blfootnote[1]{%
  \begingroup
\renewcommand\thefootnote{}\footnote{#1}%
  \addtocounter{footnote}{-1}%
  \endgroup
}
\blfootnote{All rights reserved. No reuse allowed without permission.}

\begin{abstract}

%TC:ignore
\textbf{Background:} 
From the emissions of fossil fuel power plants to the social and environmental impacts of datacenters, our current technological systems are impossible to maintain. Many of these unsustainable systems relate directly to the work of electrical and computer engineers, so it is critical for these engineers to understand the context for their work, and develop the skills and mindsets required to take action. 

\textbf{Purpose:} This paper explores the landscape of sustainability education in Electrical and Computer Engineering (ECE), to answer the following
questions: (1) How is sustainability education conceptualized and perceived in ECE departments?
(2) What pedagogical strategies and integration approaches characterize the existing landscape of sustainability education in engineering, and of ECE programs in particular? (3) What challenges limit effective sustainability education in ECE, and how might they be addressed?

\textbf{Method}: We conduct a mixed-methods study, analyzing course descriptions, program descriptions and curricula across all accredited Canadian programs, as well as accreditation materials %EJR student survey responses 
and instructor interviews.  

\textbf{Results}: We show that compared with other engineering disciplines, sustainability education in ECE is quantitatively and qualitatively less well developed, with discussions of sustainability typically limited to specific curricular structures and topics. We describe how these gaps relate to the lived experiences of %students and 
instructors, and discuss barriers that arise from both logistical issues, and more fundamentally, from narrow techno-centric perceptions of engineering identity.

\textbf{Conclusions:} There are evident gaps in the ECE curricula (both the explicit curriculum and hidden curriculum) when it comes to sustainability. We argue for diverse change strategies in order to both fill concrete curricular gaps, and to address the specific disciplinary context from which they arise.  
%TC:endignore
\end{abstract}

%\doublespacing
\section{Introduction}\label{sect:intro}
In recent years, the increasing water and energy consumption associated with widespread adoption of AI has brought the sustainability of the electronics and computing industries into the public conversation. These conversations are long overdue: beyond the environmental implications of AI, various aspects of computing have long been associated with environmentally and socially unsustainable practices. Researchers have identified nine planetary systems that provide safe, just and livable conditions for humanity, and warn that many of these systems have been perturbed past safe boundaries \citep{rockstrom2009planetary}. From the environmental harms associated with mining rare earth metals \citep{klinger2018rare} to the impacts of e-waste on ecosystems and human health \citep{grant2013health}, the computing industry impacts many of these nine systems. Research shows that the environmental impacts of computing are in fact increasing, not marginally, but measurably and dramatically: 62 million tons of e-waste were produced in 2022, a quantity projected to rise to 82 million tons by 2030, making it one of the fastest growing solid waste streams \citep{balde2024global}. Meanwhile, the International Energy Agency (\citeyear{ieaenergy}) estimates that datacenter energy utilization is projected to double by 2030, placing electricity grids under strain, and increasing construction of gas-fired power plants \citep{kollar2026fossil, gemproposed}.  

Given the potential socio-environmental impacts of their work, sustainability should be a critical concern for electrical and computer engineers. Beyond incremental changes to energy efficiency, rapid, far reaching and creative solutions are required to change the extractive paradigms that govern production, consumption and waste in computing and electronics. This requires for electrical and computer engineers to have not only the skills to design solutions that fulfill legal environmental requirements, but the ability to understand the complex social and environmental context of their work, in order to recognize and resist extractive and exploitative practices and meaningfully centre sustainability. 

We argue that given the urgency of these socio-environmental problems, sustainability education in the context of Electrical and Computer Engineering (ECE) is of critical importance. Various strategies have been proposed to teach sustainability-related competencies and mindsets in the engineering classroom, but evaluating their efficacy first requires a clear understanding of the current landscape. While prior research has outlined several frameworks defining sustainability education in engineering \citep{gutierrez2021development, gutierrez2022sustainability, brundiers2021key, wang2022sustainability, eop, wiek2011key, ram2025crafting}, a gap remains in large-scale, empirical analyses of how these theories and policies translate into practice. Research surveys in this context \citep{peters2024sustainability, thurer2018systematic, sundman2025experiential} also identify a gap in rigorous qualitative work, exploration of knowledge and value frameworks, and analysis of accreditation influences. Further, while some research discusses integration of sustainability into engineering more generally, our work shows that ECE department face unique challenges that merit focused attention. 

We address these research gaps by exploring the integration of sustainability into the ECE curriculum. From accreditation procedures and program documents that evidence different high-level approaches to sustainability education, to curriculum requirements and course descriptions that outline how sustainability competencies and topics are integrated, we map the landscape of sustainability education across 129 engineering programs at (English language) Canadian universities. Through %student surveys and long form 
interviews with educators, we discuss how this landscape shapes, and is shaped by student and educator perspectives and cultural norms. We focus primarily on the Canadian context as a case study, but also compare Canadian accreditation criteria with international counterparts.
%\newpage %Avoid splitting RQs.

Three research questions guide our analysis:
\begin{tcolorbox}[
  enhanced, breakable,
  colback=gray!5, colframe=gray!55, boxrule=0.5pt, arc=2pt,
  left=6pt, right=6pt, top=4pt, bottom=4pt,
  ]
\begin{description}[leftmargin=3.6em, style=sameline, font=\normalfont\bfseries, itemsep=3pt, parsep=0pt]
\item[RQ1] How is sustainability education conceptualized and perceived in ECE departments?
\item[RQ2] What pedagogical strategies and integration approaches characterize the existing landscape of sustainability education in engineering, and of ECE programs in particular?
\item[RQ3] What challenges limit effective sustainability education in ECE, and how might they be addressed?
\end{description}
\end{tcolorbox}

We find substantial differences between disciplines in program descriptions: none of the 19 Computer Engineering descriptions in our dataset mentions sustainability or the environment. We also distinguish the curricular settings in which sustainability-related content appears, including required disciplinary courses, electives and courses shared across engineering programs. Interviews with instructors show that some sustainability content is absent from public course descriptions, qualifying what can be inferred from document analysis alone. Interviews also contextualize and explain gaps in the curriculum, highlighting both logistical barriers, and challenges that arise from the broader departmental and disciplinary culture. Together, these findings motivate attention to both curricular provision and the place of sustainability in how ECE programs present their discipline. We conclude with a call to action, and suggest a set of concrete recommendations for educators, researchers and program administrators  (\Cref{sect:discussion}). 

%\noindent These questions and sub-questions are listed in full in \Cref{sect:study_design}, along with our study design and research methodologies. Our findings (\Cref{sect:results}) suggest that indeed, compared with other engineering departments, ECE departments face unique challenges in integrating sustainability. We observe that sustainability is unlikely to be integrated into technical or department specific coursework, and is rarely emphasized in program descriptions and promotional materials (\Cref{sect:programres}). 
%Instructor interviews and program descriptions suggest that several of these gaps relate to how students and educators understand their disciplinary identity (\Cref{sect:findings_interviews}). Given the immense social and environmental impacts of the industries ECE students work in, we argue that it is critical for  educators and institutions to address these gaps and barriers, and we propose a roadmap of concrete recommendations to do so (\Cref{sect:discussion}). 

\section{Background and Related Work}\label{sect:background}
\subsection{Defining Sustainability Education}\label{sect:defining_sus_ed}
Our research questions centre around `sustainability' education, but this term is malleable and context dependent. `Sustainability' is conceptualized in various ways by different researchers and educators. As Weisser writes in a rhetorical analysis of ‘sustainability’ in higher education, “colleges and universities use the term in a multitude of ways – often with little theoretical or ideological grounding” \citep{weisser2017defining}. In a recent systematic survey of research on sustainability in computing education, Peters et. al (\citeyear{peters2024sustainability}) find that researchers most frequently reference the 1987 Brundtland report \citep{brundtland1987our}, which calls for both intra-generational and inter-generational justice. This analysis is based on the conceptualizations of sustainability researchers however, rather than curricular materials. As Weisser (\citeyear{weisser2017defining}) points out, “almost no research has been done on the complex, shifting definitions of sustainability within US institutions of higher education”, let alone in the specific context of ECE education. 

Difficulties reaching consensus around the meaning (and value) of sustainability education also arise from fundamental differences in epistemological viewpoint and teaching philosophy. Rather than providing a fixed definition, or explicitly articulating a pedagogical philosophy, several frameworks implicitly define sustainability education by enumerating sustainability-related learning objectives. For example, based on a review of existing research, Lozano et al. (\citeyear{lozano2017connecting}) group ‘sustainability competencies’ into twelve categories, including ``personal involvement" and ``tolerance for ambiguity". Similarly, based on analysis of prior literature, and consultation with experts, Ram et al. (\citeyear{ram2025crafting}) develop a framework with 20 themes, categorized under four pillars: Environment, Social, Economic and Professional.  
Ozkan et al. (\citeyear{ozkan2026making}) argue that casting sociotechnical education as an enumeration of competencies in this way ``situate[s] it within
the institutional logics of engineering education, where learning outcomes and assessment instruments define what is
teachable and valuable". Other pedagogical perspectives centre more dynamic and subjective understandings built, for example, through reciprocal dialogue and experiential learning, which are more difficult to neatly categorize. Alternatively, educators might focus on the relationship between sustainability and epistemological justice, teaching ways of knowing instead of knowledge, and advocating for decolonizing the sustainability curriculum \citep{jimenez2023indigenizing, druker2022integration}. Approaches to sustainability education grounded in critical pedagogy \citep{freire2020pedagogy, claris2012situation} emphasize emancipation and critical theory over prescriptive learning outcomes, deconstructing power hierarchies within the educational practice to give learners agency and expose the environmental impacts of technocapitalism \citep{kahn2008education}.

Some frameworks of sustainability competencies are explicitly prescriptive, aiming to not just describe, but also influence sustainability education. For example, the Engineering One Planet (EOP) initiative has developed a framework that groups learning objectives into nine categories to provide a common language for educators and institutions \citep{eop}. Frameworks like the EOP’s, which define sustainability in order to “transform engineering practice” \citep{eop}, might be seen as examples of what Weisser (\citeyear{weisser2017defining}) describes as  “hegemonic practices [that] establish and assign a particular meaning to a word or concept, changing the reality and the material practices that emerge as a result”. %The influence of specific sustainability education definitions and frameworks like the EOP's illustrates one mechanism through which conceptualizations of sustainability education materially shape student experience and learning. 
In this work, we use the EOP framework to categorize sustainability competencies in \Cref{sect:findings_course}, but we balance this analysis strategy with a more pluralistic approach in our thematic analyses of qualitative data to avoid flattening sustainability education into a checklist that could exclude important perspectives \citep{ozkan2026making}.

\subsection{Sustainability Education Research Surveys}\label{sect:sus_ed_surveys}
Several literature reviews have been conducted on sustainability education in engineering. Peters et al.’s (\citeyear{peters2024sustainability}) recent review  provides a thorough overview of research on sustainability in computing education, including an enumeration of specific topics, pedagogical approaches, and curriculum changes. Andara et al. (\citeyear{rodriguez2018roadmapping}) and Thurer et al. (\citeyear{thurer2018systematic}) also systematically reviewed the integration of sustainability into engineering education more generally, to develop a roadmap of integration strategies. Most recently, Sundman et al. (\citeyear{sundman2025experiential}) performed a systematic review of research on experiential learning strategies for sustainability in engineering education, and Gardelle (\citeyear{gardelle2026sustainability}) surveyed European sustainability education in engineering. These literature reviews largely focus on prior academic engineering education research, rather than using primary data. Several report that the majority of studies in these areas consist of case studies and ``experience reports rather than thorough research studies" \citep{peters2024sustainability}, and call for inquiry ``taking stock of existing knowledge and synthesizing what has not yet been systematically brought together" \citep{sundman2025experiential}, something we hope to accomplish in this work.

\subsection{Cultures, Identities and Worldviews in the Engineering Department}\label{sect:culture_identity}

Throughout this paper, we discuss how perceptions of sustainability in ECE relate to disciplinary identity. Student conceptualization of engineering identity is built through explicit curriculum (eg. courses on professionalism), but also a ‘hidden curriculum’ of values and expectations embedded in institutional culture \citep{alsubaie2015hidden}.
Our discussions of this hidden curriculum of sustainability are informed by prior work on engineering ideologies and professional socialization. In particular, Cech and Sherick (\citeyear{cech2015depoliticization}) describe the ideology of depoliticization that is prominent in engineering culture: the siloing of technical and sociopolitical knowledge, and professional socialization that teaches that engineering is a primarily technical space. Through qualitative analysis, Godfrey and Parker (\citeyear{godfrey2010mapping}) map this cultural landscape, identifying six cultural dimensions. Godfrey (\citeyear{godfrey2003theoretical}) also outlines a theoretical model for cultural analysis in the context of engineering education, describing how educational artifacts and practices relate to disciplinary norms and underlying values. %Jamison et al. (\citeyear{jamison2014hybrid}) focus in particular on historical tensions between different approaches to engineering, defining three contending modes of engineering education (academic, market-driven and integrative) and describing how different teaching strategies reflect varying conceptions of engineering identity, and orientations to broader cultural contexts. 

\subsection{Change Strategies in Engineering Education}\label{sect:change_strategies}
\begin{figure}[tb!]
\centering
\includegraphics[width=\textwidth]{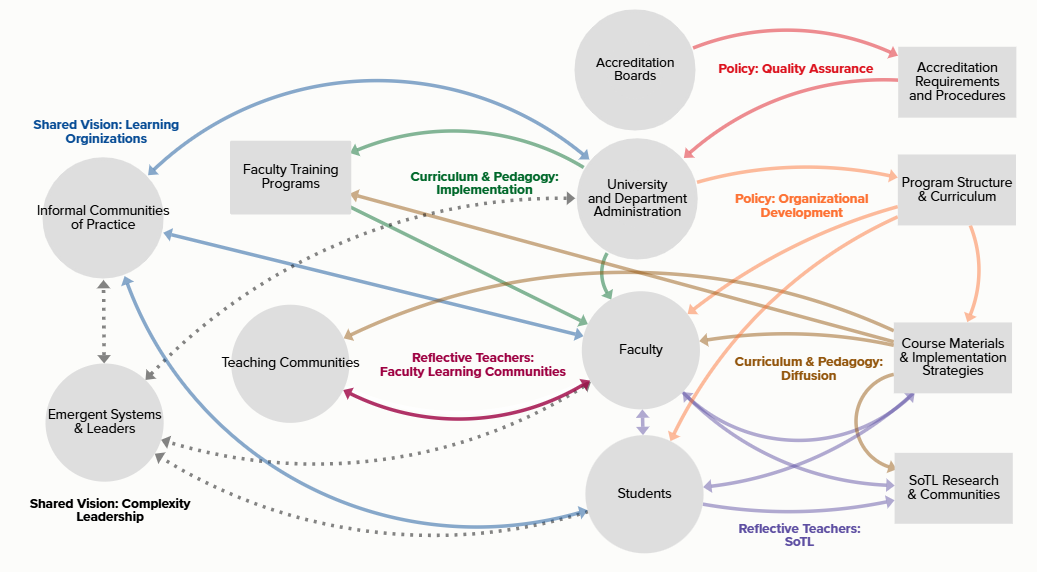}
    \caption{Concept Map of Program Structure, illustrating strategies for changing or influencing higher education. Borrego and Henderson's change strategies \citep{borrego2014increasing} are labeled.}
    \label{fig:map}
%\Description[Change strategies are represented through links in a concept map.]{Entities including Students, Faculty and Accreditation bodies influence each other and the curriculum.}
\end{figure}

Our research questions relate not only to understanding and evaluating existing sustainability education, but to actively changing this landscape. Borrego, Henderson et al. (\citeyear{henderson2011facilitating, borrego2014increasing}) identify eight change strategies for enacting change in STEM education, which we illustrate in a concept map in \Cref{fig:map}. We refer back to these change strategies throughout our recommendations in Section \ref{sect:discuss_challenges_opportunities}. 

One avenue for change is through dissemination of specific curricular and pedagogical innovations, either through faculty training or other diffusion methods. Faculty may also develop new teaching practices, either as a result of personal research and reflection, or collaboratively as part of a teaching community. As discussed in \Cref{sect:sus_ed_surveys}, much of the existing research on sustainability education has focused on the development and dissemination of such innovations. 
%TC:ignore
%TC:endignore

Change can also be enacted through more top-down policy approaches, including through accreditation.
In many countries, including Canada, engineering is a regulated profession,
and accreditation ostensibly verifies whether graduates of a given institution and program take on the disciplinary behaviours and content knowledge required by  professional licensing groups. Over the years, international agreements between regional accreditation organizations have built cooperation and coherence \citep{klassen2020global}, but accreditation organizations differ in their specific accreditation criteria and procedures. These differences have an impact. In a multiyear study of the impact of ABET criteria revisions, Prados et al. (\citeyear{prados2005quality}) report that these changes resulted in significant differences in teaching methods and learning outcomes (though Riley (\citeyear{riley2012aiding}) discusses how accreditation structures fail to achieve more systemic change). 
Byrne et al. (\citeyear{byrne2010engineering,  byrne2023evolving, byrne2013exploring}) outline sustainability-related criteria across accreditation bodies based on analysis of accreditation documents and literature, and explore the challenges associated with these requirements through workshop discussions. Staniskis and Kanlute (\citeyear{stanivskis2014engineering}) discuss the European QUESTE-SI system for accreditation of sustainability education, and its results. This work describes and discusses the different approaches used by accreditation boards to integrate sustainability education into curricula and courses, but we know of no existing research that uses primary document analysis to map the implementation of these ideas and accreditation criteria at a large scale.

In contrast with more prescriptive approaches to enacting change, Borrego and Henderson (\citeyear{borrego2014increasing}) also define ``emergent" strategies, that encourage shared vision, community and exploring other environmental features to support new forms of teaching and learning. Relevant environmental features could include, for example, the  hidden curriculum of values, norms and conceptualizations discussed in \Cref{sect:culture_identity}, and the influences of both industry ties and community-based initiatives.

\subsection{Research Gaps}\label{sect:research_gaps}
Our research questions and study design are motivated in part by  research gaps identified in prior research surveys. Ram et al. (\citeyear{ram2025crafting}) define a taxonomy to define  sustainability education in engineering and use it to evaluate engineering programs at the University of Toronto, concluding with a recommendation to extend this work geographically to other institutions. Otherwise, with the exception of [redacted], which this paper builds upon, we know of no existing research using primary document analysis to map the implementation of sustainability education on a large scale.  
The gap in such empirical studies  is emphasized in prior surveys \citep{peters2024sustainability, rodriguez2018roadmapping, thurer2018systematic, sundman2025experiential} and motivates our exploration of the existing landscape of sustainability education through program materials (RQ1 and RQ2).

Several existing surveys \citep{peters2024sustainability, rodriguez2018roadmapping, thurer2018systematic, sundman2025experiential} also emphasize the lack of  qualitative research that includes firstly, long-form interviews, and secondly, more systematic and rigorous qualitative inquiry with explicit discussion of epistemological stance, research methodology and threats to validity \citep{peters2024sustainability}. To understand the nuanced perspectives of educators, long-form interviews are a key part of this work, alongside systematic qualitative and quantitative analysis. 

\section{Study Design}\label{sect:study_design}
Our purpose is to explore the existing landscape of sustainability education in ECE, and to identify strategies for shaping sustainability education for the better. We list our research questions in full below, alongside relevant results and discussion sections (italicized). 

\begin{tcolorbox}[
  enhanced, breakable,
  colback=gray!5, colframe=gray!55, boxrule=0.5pt, arc=2pt,
  title=Research Questions, fonttitle=\bfseries,
  coltitle=black, colbacktitle=gray!20,
  left=6pt, right=6pt, top=4pt, bottom=4pt,
  ]
\begin{description}[leftmargin=3.6em, style=sameline, font=\normalfont\bfseries, itemsep=2pt, parsep=0pt]
\item[RQ1] How is sustainability education conceptualized and perceived in ECE departments? \textit{(Section \ref{sect:discuss_conceptualization})}
  \begin{description}[leftmargin=2.6em, style=sameline, font=\normalfont\bfseries, itemsep=1pt, topsep=2pt]
  \item[RQ1.1] How do ECE educators 
  evaluate the quality or   `substantivity' of  sustainability education? \textit{(Section \ref{sect:findings_interviews})}
  \item[RQ1.2] How is sustainability framed in the `hidden curriculum' of engineering departments, in particular program documentation? \textit{(Section \ref{sect:findings_program})}
  \end{description}
\item[RQ2] What pedagogical strategies and integration approaches characterize the existing landscape of sustainability education in engineering, and of ECE programs in particular? \textit{(Section \ref{sect:discuss_landscape})}
  \begin{description}[leftmargin=2.6em, style=sameline, font=\normalfont\bfseries, itemsep=1pt, topsep=2pt]
  \item[RQ2.1] How are topics and competencies related to sustainability currently integrated into engineering programs? \textit{(Section \ref{sect:findings_accred}, \ref{sect:findings_curriculum}, \ref{sect:findings_course})}
  \item[RQ2.2] How do the strategies used in ECE compare with other engineering disciplines? \textit{(Section \ref{sect:findings_curriculum}, \ref{sect:findings_course})}
  \end{description}
\item[RQ3] What challenges limit effective sustainability education in ECE, and how might they be addressed? \textit{(Section \ref{sect:discuss_challenges_opportunities})}
  \begin{description}[leftmargin=2.6em, style=sameline, font=\normalfont\bfseries, itemsep=1pt, topsep=2pt]
  \item[RQ3.1] What barriers and conflicts prevent more effective or widespread sustainability education in ECE? \textit{(see all results and discussions)}
  \item[RQ3.2] What opportunities exist for addressing these barriers, and increasing integration sustainability in ECE education? \textit{(see all results and discussions)}
  \end{description}
\end{description}
\end{tcolorbox}

\subsection{Theoretical Perspectives and Approach}\label{sect:methodology}
\subsubsection{Research Paradigm and Approach}
Pragmatism is the underlying research paradigm in this work. Our perspective is that given the ongoing and escalating climate crisis, sustainability education urgently needs to be improved, increased and diversified. 
As we discuss in \Cref{sect:discuss_conceptualization}, we find that many of the challenges associated with teaching sustainability in engineering relate to conflicts in epistemological perspectives and conceptualizations of sustainability and engineering between instructors and students.
Given the diversity of opinions and perspectives on sustainability education, we adopt a pragmatic approach and aim to provide pluralistic discussions that embrace oppositional viewpoints and facilitate coordinated action.

We also note however, that our interpretations and discussions (in particular \Cref{sect:discuss_conceptualization}) are shaped by our own personal perspectives on education (as described in \Cref{sect:positionality}). In particular, these perspectives are informed by critical pedagogy, an approach to education originally described by Paulo Friere (\citeyear{freire2020pedagogy}), and later applied in the context of engineering by various researchers \citep{mejia2018critical, riley2003pedagogies, claris2012situation}). 

\subsubsection{Research Methodology and Methods}
In line with our pragmatic approach, we employ mixed methods to collect and analyze a combination of quantitative and qualitative data. We include large scale quantitative analysis to empirically evaluate broad patterns in sustainability education across universities. We also include interviews 
and qualitative analysis of program and course descriptions to contextualize quantitative findings with more nuance, and to include several different perspectives on these topics. To strengthen the validity and quality of our research, we refer in particular to the ACM SIGSOFT Empirical Standards \citep{ralph2020empirical}, Mays and Pope's (\citeyear{mays2000assessing}) guidelines on quality in qualitative research, and Gillespie et al.'s (\citeyear{gillespie2024pragmatism}) guidebook on pragmatism and mixed methods. We discuss threats to validity in \Cref{sect:discuss_validity}.

\subsection{Data Collection and Analysis}\label{sect:data_collection} 
%We used different strategies to analyze and collect various types of data, in some cases (for program descriptions for example) employing multi-resolution analysis to analyze the same data using different approaches. 

\subsubsection{Accreditation Documents and Criteria}\label{sect:accred_analysis} 
Canadian accreditation requirements, history, assessment procedures, and documents relating to sustainability education were sourced from the Engineers Canada website. Since Engineers Canada accreditation procedures are influenced by their membership in International Engineering Alliance (IEA) agreements, we collected similar documentation from the IEA website. %, and from ABET, as a point of comparison. 
We also collected accreditation criteria from all full signatories of the IEA's Washington Accord, of which there are 25. A clear English-language description of accreditation criteria was available on the public facing website of each signatory (URLs are listed in Appendix \ref{app:appendix_signatories}). 

\subsubsection{Program Descriptions, Curricula and Course Descriptions}\label{sect:doc_analysis} 
We manually collected public curricula, course and program data from all Canadian English-language engineering programs for the six most common engineering disciplines: Chemical (CHEM), Civil (CIV), Environmental (ENV), Mechanical (MECH), Computer (COMP) and Electrical (ELEC). This data collection was performed for 129 programs in total, in November 2025. We include the full list of programs and program description URLs in Appendix \ref{app:appendix_programs}: 

\begin{itemize}[listparindent=\parindent]
\item\textbf{Program Descriptions:} %For each undergraduate program, we collected 
All text from the primary webpage introducing the program to prospective students. 

For the large majority of programs, prospective undergraduate students were directed to a single page that provided a program introduction and overview. When similar information was published across multiple sites, choosing a single webpage to use in our analysis involved a degree of subjectivity. In these cases, we selected descriptions that were explicitly aimed at prospective students, with a preference for program descriptions published on the departmental website (as opposed to by university admissions).

To analyze sustainability-related discourses in program promotional materials, we 
began with preliminary data familiarization, followed by systematic keyword coding using NVivo, through which we generated a codebook of key data features, with particular attention to any keywords related to the environment or sustainability. This keyword coding forms the basis of the quantitative results in Section \ref{sect:findings_program} and Table \ref{tab_descriptions}. Based on this keyword coding, we iteratively developed and grouped patterns and ideas that were common across multiple program descriptions. In this stage, we employed inductive thematic analysis to outline the ways in which sustainability was referenced in these documents.
 Finally, we performed an additional pass through the program descriptions of ECE programs specifically, to more carefully evaluate how those programs in particular related to our inductively developed conceptual models and to find illustrative examples.
\item\textbf{Program curricula:} %For each program, we recorded 
Program requirements, including required courses, elective courses and program options for the most recent cohort. 

This information was typically available through either the departmental program description, or course calendar. If multiple curricula were described for different academic cohorts, we chose the most recent version (although in many cases this information was unspecified). To analyze curricula, we parsed public facing program documentation and recorded all sustainability-related program requirements in a database, allowing us to reason quantitatively about how sustainability-related courses were integrated into programs.
\item\textbf{Course descriptions:} Descriptions of each program’s courses, compiled into a database. 

In this database, we included all optional electives offered by the corresponding engineering departments. Courses offered by faculties and departments outside of engineering were only included when completion of those courses was a program requirement. We then used AI-assisted keyword coding and categorization to analyze all course descriptions. With 774 sustainability-related courses (and many more which did not relate to sustainability), it was not possible to manually analyze all course descriptions. Instead, we manually analyzed a subset of fifty course descriptions to familiarize ourselves with how sustainability competencies were typically described, and to develop a codebook of common keywords and phrases. Course descriptions were analyzed deductively using the pre-existing EOP framework of sustainability competencies \citep{eop}, rather than through an inductive process. We outlined our categorization criteria, including examples, keywords and phrases, and corner-cases, and used Claude LLM (claude-sonnet-4.6) to categorize the remainder of the courses according to our criteria, based on which EOP competencies (up to two) they related to most closely. Finally, we manually analyzed another subset of fifty course descriptions and calculated Cohen’s Kappa metric \citep{mchugh2012interrater} to evaluate how well the AI-based analysis aligned with human analysis.
\end{itemize} 
 
 \subsubsection{Instructor Interviews} 
 In semi-structured interviews, we asked six instructors about their experiences integrating concepts and learning objectives related to sustainability into their courses. Each instructor taught at least one engineering course in a program listed in Appendix \ref{app:appendix_programs}, and five of the six interviewees had already integrated discussions of sustainability into their courses. Eligible faculty interviewees were identified through personal connections
and networks, and were contacted directly by email. Each participant was asked a consistent set of interview questions that touched on our broad areas of inquiry, but our semi-structured interview approach allowed for follow-up probing and discussion of emergent ideas. Interviews were conducted in person or online by the first author and lasted 20 to 60 minutes.
 
 Interview transcripts were analyzed using thematic analysis, using a similar approach described in \Cref{sect:doc_analysis}, but with a focus on more qualitative inductive theme development, rather than on systematic keyword coding. Compared with our more quantitative characterization of the sustainability education landscape, using interviews we aimed to explore underlying student and educator conceptualizations of sustainability and engineering identity, necessitating a more reflexive and qualitative approach, informed by Terry et al.'s (\citeyear{terry2017thematic})  guidelines on reflexive thematic analysis.

\section{Results}\label{sect:results}
We organize our results according to data type, beginning with our analysis of accreditation documents and procedures (\Cref{sect:findings_accred}), followed by program descriptions, curricula and course descriptions (\Cref{sect:programs})
 and our instructor interviews (\Cref{sect:findings_interviews}). 

\subsection{Accreditation Documents and Criteria}\label{sect:findings_accred}
In Canada,  Engineers Canada performs accreditation, and as a signatory of the Washington Accord, is committed to doing so in a way consistent with the International Engineering Alliance (IEA)'s guidelines. The IEA's Graduate Attribute and Professional Competencies (GAPC) framework outlines 11 `graduate attributes', and Washington accord signatories are required to publicly document and uphold a set of substantially equivalent educational outcomes. There are notable differences however in how individual accreditation bodies adapt these attributes in their own jurisdictions when it comes to sustainability. We analyze these criteria and compare them across all full signatories (listed in Appendix \ref{app:appendix_signatories}) 
to understand how the Canadian criteria and position compare with international equivalents. 

Several signatories adopt Version 3 (PEC and CAST) or  Version 4 (CACEI, EngNZ, IES, IESL, NBA and IEB) of the GAPC attributes either exactly or with minor wording changes. EA (and to a lesser extent, ECUK and ABET) significantly restructure their accreditation outcomes in a way that makes direct comparison difficult. The remainder adopt a similar structure to GAPC 3.0 or 4.0, but adapt, add or combine attributes. For these signatories, Table \ref{tab:graduate_attributes} summarizes which graduate attributes refer to environmental conservation or sustainability. 

Three of the GAPC 4.0 attributes discuss the environment or sustainable development:
\begin{itemize}
    \item \textbf{Problem Analysis}: ``Identify, formulate, research literature and analyze complex engineering problems reaching substantiated conclusions using first principles of mathematics, natural sciences and engineering sciences \textbf{\textit{with holistic considerations for sustainable development}}."

    The highlighted text was added in GAPC 4.0, and many signatories, including Engineers Canada and ABET have not included this addition. 
    \item \textbf{Design}: ``Design creative solutions for complex engineering problems and design systems, components or processes to meet identified needs with appropriate consideration for public health and safety, whole-life cost, \textbf{\textit{net zero carbon}} as well as resource, cultural, societal, and \textbf{\textit{environmental considerations}} as required".
    
    Some signatories' criteria similarly refer to environmental `considerations', `matters', `factors', `aspects', `constraints', or `ecological limitations'.% In other cases, design attributes do not discuss any such considerations, or briefly refer to `realistic constraints' without explicitly referring to sustainability. 
    \item \textbf{The Engineer and the World}: ``When solving complex engineering problems, analyze and evaluate \textbf{\textit{sustainable development impacts}} to: society, the economy, \textbf{\textit{sustainability}}, health and safety, legal frameworks, and \textbf{\textit{the environment.}}"
    
    For some signatories, and in GAPC 3.0, social and environmental impacts are split into two separate attributes. In other cases (ABET, EI and IABEE) these impacts are discussed instead in an attribute centred on ethics or professional responsibility. %, requiring not only understanding of social and environmental impacts, but also for engineers to take make ethical judgments based on those impacts. 
\end{itemize}  
%TC:ignore
\begin{table}[h]
\begin{threeparttable}
\caption{Reference to sustainability or environmental considerations across Graduate Attributes}
\small
\label{tab:graduate_attributes}
\centering
  \begin{tabular}{>{\centering\arraybackslash}m{4cm}| p{.2cm} p{.2cm} p{.2cm} p{.2cm} p{.2cm}p{.2cm}p{.2cm}p{.2cm}p{.2cm}p{.2cm}p{.2cm}p{.2cm}p{.2cm}p{.2cm}p{.2cm}p{.2cm}p{.2cm}p{.2cm}p{.2cm} }
    \hline
    Graduate Attribute & \rotatebox[origin=c]{60}{GAPC 4.0$\dagger$} & \rotatebox[origin=c]{60}{GAPC 3.0$\ddagger$} & \rotatebox[origin=c]{60}{ABET} & \rotatebox[origin=c]{60}{ABEEK} & \rotatebox[origin=c]{60}{AEER} & \rotatebox[origin=c]{60}{MUDEK} & \rotatebox[origin=c]{60}{BEM} & \rotatebox[origin=c]{60}{CFIA} &  \rotatebox[origin=c]{60}{ECSA} & \rotatebox[origin=c]{60}{ECUK} & \rotatebox[origin=c]{60}{EC} & \rotatebox[origin=c]{60}{EI} & \rotatebox[origin=c]{60}{IABEE} & \rotatebox[origin=c]{60}{IEET} & \rotatebox[origin=c]{60}{ICACIT} & \rotatebox[origin=c]{60}{JABEE} & \rotatebox[origin=c]{60}{PTC} & \rotatebox[origin=c]{60}{HKIE} \\ 
    \hline
    Problem Analysis &
      $\checkmark$ &  & & $\checkmark$ & & 
      $\checkmark$ & &  &$\checkmark$ 
      & & &  $\checkmark$ & & &
      $\checkmark$ & & &\\
    Design / Development of Solutions &
      $\checkmark$& $\checkmark$ &  $\checkmark$ & $\checkmark$ & $\checkmark$ &
      & $\checkmark$ & $\checkmark$&  
      $\checkmark$ &$\checkmark$&$\checkmark$  &$\checkmark$ & $\checkmark$&&
      $\checkmark$ &$\checkmark$ &$\checkmark$ &$\checkmark$ \\
    `The Engineer and the World' or `Environment and Sustainability' &
    $\checkmark$ & $\checkmark$ & * & $\checkmark$ & 
    $\checkmark$ & $\checkmark$  & $\checkmark$ &$\checkmark$ &
    $\checkmark$ &$\checkmark$ &$\checkmark$&  * & * & $\checkmark$ & 
    $\checkmark$ & $\checkmark$ &$\checkmark$ & $\checkmark$  \\
    Ethics &
    && $\checkmark$ &&
    &&&&
    &&&$\checkmark$&$\checkmark$&
    &&& \\
    \hline
  \end{tabular}
    \begin{tablenotes}
      \item[$\dagger$] and CACEI, EngNZ, IES, IESL, NBA and IEB;
      \item[$\ddagger$] and PEC and CAST;
      \item[*] This signatory does not have a directly equivalent attribute to compare against. 
    \end{tablenotes}
  \end{threeparttable}
\end{table}
%TC:endignore
\subsection{Program and Course Implementation}\label{sect:programs}
Curriculum documentation, including program and course descriptions, and program requirements, outline the existing sustainability education that engineers receive throughout their undergraduate degrees. These are not only administrative requirements, but cultural artifacts that  reveal how institutions and disciplines conceptualize the engineering profession. 

\subsubsection{Program Descriptions and Rhetoric}\label{sect:findings_program}
Program descriptions include a wide range of information, including descriptions of facilities, application information, student testimonials and various other descriptions, lists and statistics. We group this information into five categories: Call to Action, Introduction to the Discipline, Careers / Future Work, Learning Objectives and Program Features. Not all categories were present in all program descriptions, and many program descriptions also included additional information (for example, specific application instructions). 
\begin{enumerate}
    \item 
\textbf{Call to Action}: Many descriptions began with a brief motivational example, problem definition or action-oriented phrase. The call to action most often related to solving social or environmental problems (“Ready to help solve some of the world’s most pressing energy, climate change and sustainability challenges?” – University of Waterloo, ENV), described career opportunities (“Power up your career success and be on the forefront of innovation” – UWindsor, ELEC) or underscored the importance of the field in everyday life. 
\item \textbf{Disciplinary Identity}: Most program descriptions described the engineering discipline and the role of engineers in a discipline-specific context, whether in more practical terms (“The growing field of Computer Engineering helps develop systems and solutions for the technological industry” – UofG, COMP), or more metaphorically (“Chemical Engineers are the alchemists of our world…” – TMU, CHEM).
\item \textbf{Future Work}: Program descriptions either broadly describe career paths in different sectors (“Graduates compete for top engineering jobs in the private sector, within governmental organizations and with many not-for-profit organizations” – UofR, ELEC), enumerate specific fields (“Graduates from this program work in various industries, including aerospace, automotive, green energy, and oil” – University of Waterloo, MECH), list specific positions and companies, 
or outline possible research areas. They also frequently highlight the achievements and employability of previous graduates. 
\item \textbf{Learning Objectives and Pedagogy}: Most program descriptions described course content or learning objectives (“you will learn essential skills in electronics, circuit analysis and electromagnetics” – UBC, ELEC), approaches (“U of R teaches a systems approach that solves engineering problems through more holistic thinking” – UofR, ELEC) and pedagogical strategies. In particular, program descriptions often highlighted experiential learning opportunities, including labs, project-based courses and co-op opportunities. 
\item \textbf{Program Features and Descriptions}: Many descriptions also described various extracurricular opportunities including paid work experience opportunities, student groups and clubs and study abroad options. They also highlighted key features of the university or program, including industry connections, 
student experience (“you’ll learn by being immersed in a structured, highly demanding environment” – BCIT, CIV), rankings, reputation, accreditation status and high level university values (“core values on our campus [...] embody diversity equity and inclusion” – UofR, ELEC).
\end{enumerate}

%TC:ignore
\begin{table}
\small
  \caption{Sustainability Patterns in Program Descriptions}
  \label{tab:sus_patterns}
  \begin{tabular}{>{\centering\arraybackslash}m{4.3cm}||p{5cm}|p{5cm}}
     Pattern & Description & Example \\ \hline \hline
    Sustainability as \mbox{Potential} Application & Sustainability-related technologies or products are listed as potential applications of engineering knowledge or skills, listed alongside other applications or fields (often including traditionally extractive industries). & \textit{“They work in any number of industries and during their careers, they may face a variety of challenges, such as optimizing processes and monitoring pollution, [...]”} 
- uOttawa, CHEM \\ \hline
    Sustainability as \mbox{Motivation} & Sustainability,  environmental protection, or environmental concerns are used to emphasize the importance or urgency of the work of engineers in the discipline. & \textit{“Are you concerned about human welfare and the environment?”} – UBC, MECH \\ \hline
    Sustainability as Core Engineering Principle & Sustainability is positioned as a key part of the role of the engineer. & \textit{“Civil engineers are [...] stewards of the natural environment”} – BCIT, CIV \\ \hline
    Sustainability as Design Metric or Regulation & Sustainability (or limiting environmental harm) is framed as a regulatory requirement or one of many design criteria.  & \textit{“As a chemical engineer you might help industry to meet their sustainability targets”} - UNB, CHEM \\ \hline
    Sustainability Absent & Many programs include no mention of sustainability or the environment. &  \\ \hline
  
\end{tabular}
\end{table}
%TC:endignore
 For each section of the program description, we considered in particular how sustainability and related topics were framed, and identified five main patterns, listed in Table \ref{tab:sus_patterns}. There are noticeable differences between disciplines in how often each pattern is used. Across all 19 accredited Canadian English-language Computer Engineering program descriptions, none mentioned sustainability, the environment, or explicitly sustainable technologies or applications. Of 29 Electrical Engineering programs, 41\% did not discuss sustainability. 38\% listed renewable energy generation as a possible application of electrical engineering. 
An additional 10\% discussed  renewable energy more centrally, as a potential motivation to study electrical engineering:
\begin{quote}
``The world is looking to develop sustainable, environmentally friendly and diversified sources of electrical energy." - Western, ELEC
\end{quote}
Finally, 10\% of Electrical Engineering programs mentioned consideration of sustainability or efficiency during the design process:
\begin{quote} ``Use a systems approach that focuses on economic, social, environmental, safety and administrative considerations relevant to the daily work of engineering." - UofR, ELEC \end{quote} 
This is significantly less explicit discussion of sustainability or environmental conservation than other engineering disciplines' program descriptions, most of which at least mention sustainability, often describing it as a central aspect of engineering. Table \ref{tab_descriptions} lists the number of program descriptions in which sustainability is discussed in each discipline (Patterns 1-4).
%TC:ignore
\begin{table}[h]
\caption{References to sustainability in program descriptions by discipline}
\label{tab_descriptions}
\centering
  \begin{tabular}{ l l p{5cm} }
    \hline
    Discipline & \# Programs & Sustainability in Program Descriptions (Patterns 1-4) \\ 
    \hline
    Computer & 19 & 0 (0\%)  \\
    Electrical & 29 & 17 (58\%) \\
    Civil & 26 & 17 (65\%) \\
    Mechanical & 29 & 20 (69\%) \\
    Chemical & 17 & 16 (94\%) \\
    Environmental & 9 & 9 (100\%) \\
    \hline
  \end{tabular}
\end{table}
%TC:endignore
Instead of describing positive impacts on the environment or society, many ECE program descriptions focus on the number of jobs available to graduates, and the ubiquity of these technologies in modern life (or, implicitly, modern life in the Global North). 
In Computer Engineering in particular, 20\% of program descriptions mention technologies that are ``shaping our world" or ``shape modern life", while in Electrical Engineering, 28\% of program descriptions describe how the discipline ``powers our future/modern life/our world". 

Program descriptions certainly vary in terms of length and tone between universities, but there is a consistent difference in how sustainability is addressed between disciplines within the same university. For example, for the 18 universities that have both Computer Engineering and Civil Engineering programs, none of the Computer Engineering program descriptions and 11 Civil Engineering program descriptions mentioned sustainability or the environment. 

We note that program descriptions may not always be representative of the degree to which topics are discussed throughout the programs themselves, and are not a measure of actual teaching or accreditation (particularly since all programs are subject to the same accreditation requirements).  

\subsubsection{Program Structures and Curricula}\label{sect:findings_curriculum}
\begin{figure}[htbp]
    \centering
    \includegraphics[width=\columnwidth]{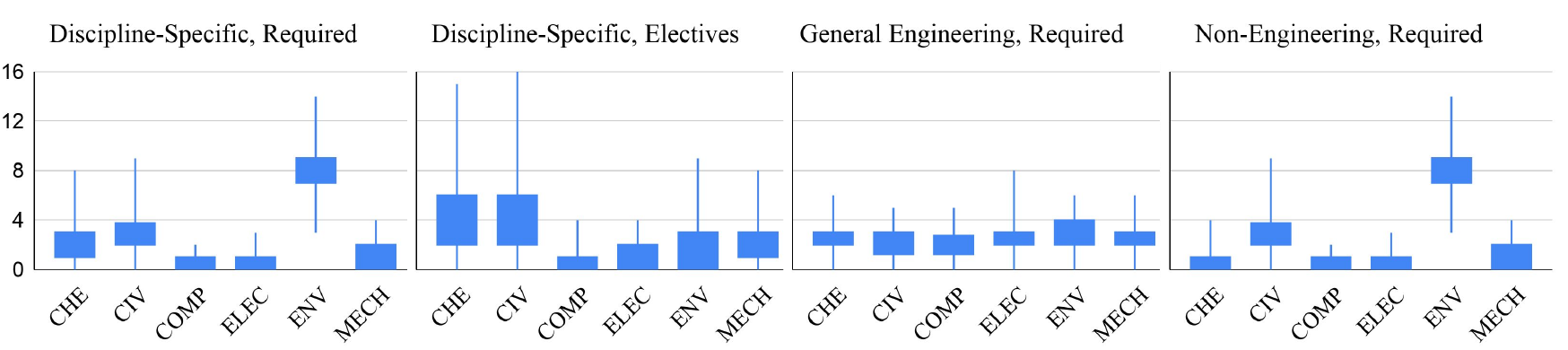}
    \caption{Sustainability integration by course type across disciplines (uses per program)}
    \label{fig:curricula_fig}
    %\Description[A box and whisker chart shows the number of sustainability courses per program.]{Course count is separated by discipline. Major trends are described in the text.}
\end{figure}

Through an analysis of curriculum structures across 129 engineering programs, we identified four structural strategies for sustainability coverage. 
A single program can (and often does) combine multiple strategies. 
\begin{enumerate}
\item  \textbf{Discipline-Specific, Required Courses}:
Integration of sustainability related topics or learning objectives across required discipline-specific technical or design courses. 
\item	\textbf{Discipline-Specific Electives}:
Optional, discipline-specific technical electives that relate to sustainability. In some cases (particularly Civil Engineering), a collection of departmental and non-departmental electives relating to sustainability were grouped into a sustainability ‘stream’.
\item \textbf{General Engineering Required Courses}: Integration of sustainability through  extra-departmental or general engineering courses. For example, in many programs the impacts of engineering on society and the environment are taught through an “Introduction to the Engineering Profession” course (or similar) that is shared by all engineering programs. 
\item \textbf{Non-Engineering Required Courses}:
Students are required to take a specific sustainability-related course, or to choose one elective out of a selection of sustainability-related electives offered by different faculties (eg. Sociology).

 \end{enumerate}
We tracked how often programs employed each strategy to highlight dominant curricular patterns across the country. In Figure \ref{fig:curricula_fig}, we list the average number of times each integration strategy is used (ie. for elective-based integration, the number of optional sustainability-related electives available in the department) across disciplines. The use of different integration approaches is highly discipline dependent. For example, in Civil and Environmental Engineering, sustainability is likely to be discussed in core technical coursework, while in Environmental Engineering, students are often required to take sustainability-related courses from outside of engineering. 

In around half of ECE programs, a required ECE course discusses sustainability or the environment. These courses are often either project-based design courses or courses centred around professional practice. For example, ELEC290 at Queen's University:
\begin{quote}\textbf{Electrical and Computer Engineering Design and Practice}: This course encompasses team-based design to solve complex open-ended problems. Instruction is provided on problem definition, creation of ideas, and decision making that considers technical, economic, \textbf{\textit{societal, and environmental factors}} [...]
\end{quote}
On average, ECE students are also required to take one additional general engineering course in which sustainability is discussed. These courses are often introductions to the engineering profession, like ENGG160 at the University of Alberta:
\begin{quote}\textbf{Introduction to Engineering Design, Communication, and Profession}: Fundamental design process and theory in a multidisciplinary context. Importance, in engineering design, of communications; team work; the engineering disciplines, career fields; professional responsibilities of the engineer including \textbf{\textit{elements of ethics, equity, concepts of sustainable development and environmental stewardship}} [...] 
\end{quote}
Optional electives offered by the ECE department that involve sustainability are rare, and most frequently center around power electronics or energy systems, like EE448 at the University of Saskatchewan:
\begin{quote}
\textbf{\textit{Renewable Energy and Power Systems}}: Covers salient-pole synchronous machines; solar photovoltaic (PV) power technology [...] renewable energy integration and microgrids; and \textit{\textbf{various renewable energy conversion systems}}.
\end{quote}
Finally, several ECE programs require students to take sustainability-related complementary studies courses offered by other faculties, like ANTH2430 at the University of Manitoba:
\begin{quote}
\textbf{\textit{Ecology, Technology and Society}}
Ecological analysis of the interplay of socio-political and technological processes in different types of societies. Focus upon the \textbf{\textit{ecological side-effects and selected technologies}}, economic mechanisms and political institutions.
\end{quote}
%TC:ignore

\begin{figure}[htbp]
  \begin{subfigure}{.5\textwidth}
  \centering
    \includegraphics[width=1\linewidth]{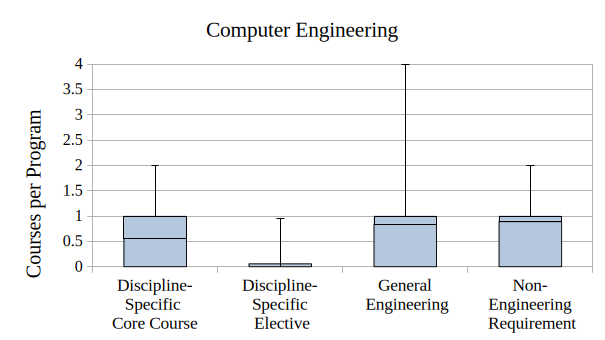}
  \end{subfigure}%
  \begin{subfigure}{.5\textwidth}
  \centering
    \includegraphics[width=1\linewidth]{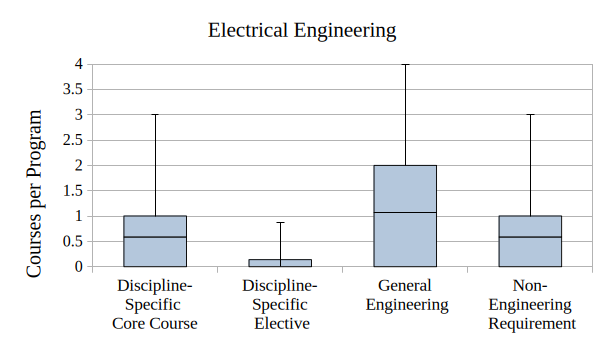}
  \end{subfigure}
  \caption{Sustainability integration strategies in Electrical and Computer Engineering, at the curricular level}
    \label{fig:curricula_ece}
    %\Description[A box and whisker chart shows the different types of sustainability courses available in Computer and Electrical Engineering.]{There are particularly few discipline-specific ECE electives that reference sustainability.}
\end{figure}
%TC:endignore
\subsubsection{Course Content}\label{sect:findings_course}
Analysis of individual course descriptions shows not only \textit{how} and \textit{when} students are developing sustainability-related competencies, but also \textit{which} sustainability competencies are being developed. As outlined in Section 
 \ref{sect:doc_analysis}, we used an AI-assisted process to identify all engineering course descriptions that touch on sustainability, and to categorize them based on the EOP competencies they related to. For example, the following course (Dalhousie’s CHEE4900) was categorized as Environmental Impact Assessment and Material Selection:

\begin{quote} ``This course will introduce methods for choosing materials, and the processes used to produce them, for mechanical design. [...] Cost, \textbf{\textit{environmental impact, life cycle and sustainable materials/processes}} will be included [...]"
\end{quote}

To assess the reliability of our AI-based analysis, we calculated inter-rater reliability using a sample of 50 courses categorized both by AI and manually. Cohen’s Kappa ($\kappa$) was .74, which confirms that while there was substantial agreement between our subjective categorization and the AI model, the process certainly involved a degree of subjectivity and ambiguity. Cohen's Kappa is listed by EOP category in Table \ref{tab:cohens_kappa}. The poorest agreement is for categories ‘Design’ and ‘Communication and Teamwork’: typically disagreements occurred when AI labelled courses as falling into one of these categories when manual categorization did not. For example, the following course was categorized as “Environmental Impact” manually, and both “Environmental Impact” and “Design” by the AI model: \begin{quote} “Combustion systems are an essential part of our society, however, they are also the dominant source of atmospheric pollutants. This course will focus on investigation of combustion systems for the purpose of reducing atmospheric emissions.” - ENGG4820, UofG \end{quote}

\begin{table}[h]
\begin{center}
\begin{threeparttable}
\caption{Cohens Kappa (interrater reliability) by EOP category}
\small
\label{tab:cohens_kappa}
\centering
  \begin{tabular}{p{7cm} p{3cm} }
    \hline
    IEA Graduate Attribute & Cohen's Kappa \\ 
    \hline
    Systems Thinking	& .65 \\
    Environmental Literacy	& .75\\
    Responsible Business and Economy	& 1\\
    Social Responsibility	& .90 \\
    Environmental Impact Assessment	& .63 \\
    Material Selection	& 1 \\
    Design & .61 \\
    Critical Thinking	& .70 \\
    Communication and Teamwork & .56 \\
    \hline
  \end{tabular}
  \end{threeparttable}
  \end{center}
\end{table}

\begin{figure}[htbp]
    \centering
    \includegraphics[width=\columnwidth]{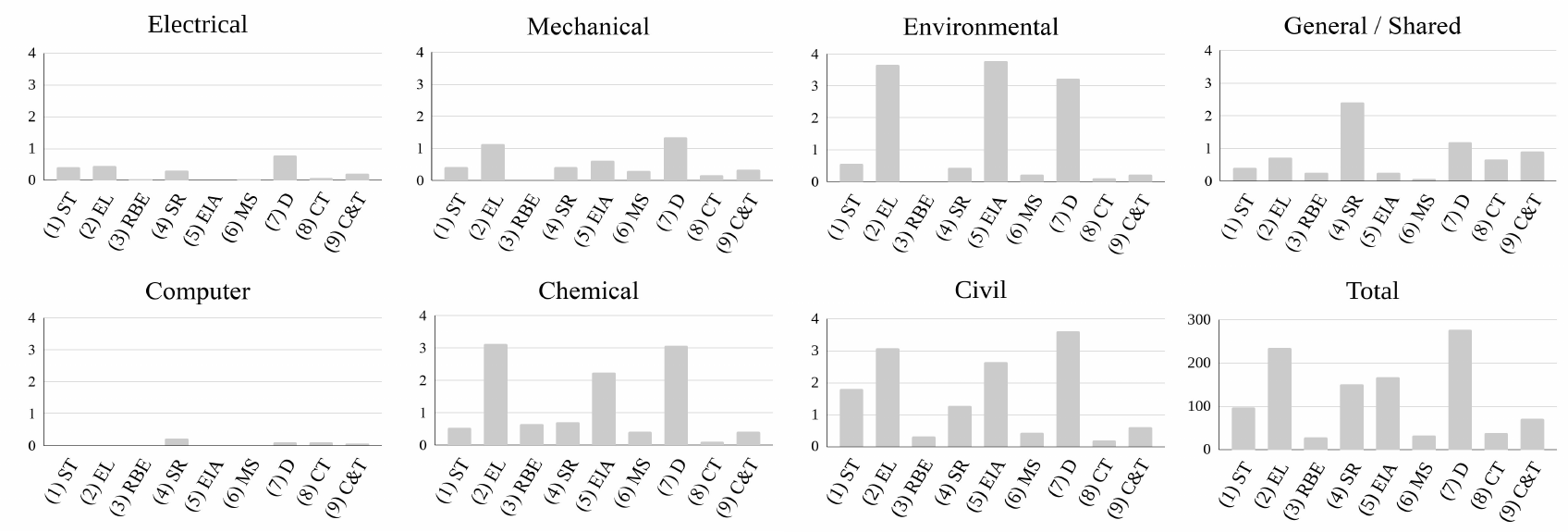}
    \caption{Total courses (per university) covering EOP sustainability competencies, separated based on the department offering them (1. Systems Thinking, 2. Environmental Literacy, 3.Responsible Business and Economy, 4. Social Responsibility, 5. Environmental Impact Assessment, 6. Material Selection, 7. Design, 8. Critical Thinking, 9. Communication and Teamwork)}
    \label{fig:course_fig}
    %\Description[A bar chart shows the number of courses per department, categorized by EOP sustainability competency.]{Computer Engineering has the fewest across all EOP competencies, followed by Electrical and Mechanical Engineering. Environmental Literacy, Social Responsibility, Environmental Impact Assessment and Design are the best represented categories.}
\end{figure}

Categorization results are summarized in Figure \ref{fig:course_fig}, with courses grouped based on the department offering them rather than by academic program, since students in a given academic program may take courses from several different departments. `General/Shared' courses are offered by the engineering faculty, but are not associated with a specific department. 
%TC:ignore
%TC:endignore

Computer Engineering departments offer far fewer courses with descriptions referring to sustainability than any other department. We found no Computer Engineering courses that related to sustainable material selection, environmental impact assessment, sustainable business practices, environmental literacy or systems thinking (in the context of sustainability). Those that did relate to sustainability mentioned ethical decision making or `ethical issues' without discussing sustainability or the environment specifically. Electrical Engineering departments offered the second fewest sustainability-related courses, with courses most frequently discussing sustainability in the context of design, mentioning `environmental considerations' in a more general context, or focusing specifically on the design of renewable energy systems or electric vehicles. This is consistent with Ram et al's (\citeyear{ram2025crafting}) analysis of the University of Toronto's engineering programs, which similarly shows the least sustainability coverage in the ECE department. 

It would be ideal to have instructors, who fully understand the course content, context and learning objectives, categorize their own courses. Course descriptions are not necessarily fully descriptive of content or up to date, and are often slightly ambiguous or open to multiple interpretations. For this reason, we supplement this analysis of course descriptions with instructor interviews, to better understand how concepts are actually integrated into real courses, and to evaluate the extent to which course descriptions are representative of actual course content (\Cref{sect:findings_interviews}).

\subsection{Instructor Experiences}\label{sect:findings_interviews}

We interviewed six engineering instructors teaching topics related to sustainability: Maria, Arjun, Chris, Andrew, Matt and Amir. These names are pseudonyms, and in the results below, potentially identifying details have been removed or anonymized.
\subsubsection{Instructor and Course Information}
Of the six instructors interviewed, four taught ECE courses, and two taught Civil or Chemical Engineering courses. From energy systems to ethical reasoning, the competencies and content they taught varied significantly. Most instructors had integrated some level of discussion of sustainability into at least two different undergraduate courses. Arjun was the only exception; he had briefly discussed a topic related to sustainability in one course, but had not otherwise taught sustainability related competencies. Matt, Chris and Andrew's courses included a Capstone design course, and Chris and Matt also taught courses specifically centred on social impacts or socio-environmental contexts of engineering. All instructors had also taught one or more upper-year technical courses. 

Several of the courses the instructors described had not been included in our earlier analysis of course descriptions because their course descriptions did not explicitly mention sustainability or associated topics. As Andrew describes for example, his course's relationship to sustainability ``probably doesn't come through necessarily in the course descriptions because it's kind of integrated throughout all of the material". Amir described how course descriptions could also easily become out of date: 
\begin{quote}``To be frank, I
never went back and updated [my course descriptions]. It's just something that exists on the website. It's not something that I can just go edit. I have to send a request, someone has to edit it..."
\end{quote}

\subsubsection{Instructor Motivations}
Instructors described interrelated motivations for integrating new discussions of sustainable development or sustainable technologies into their courses:
\begin{itemize}
    \item \textbf{Driving pro-social and environmental changes:} Instructors viewed sustainability education as an opportunity to enact pro-social or environmental change. For example, Amir described changing industry perspectives:
    \begin{quote}``We want more people in industry to also be thinking of these things. That's part of the reason I think it's worth exposing our students to these questions, because they can go in the industry [...] and have these questions in the back of their minds."
    \end{quote}
    Maria described how sustainability and prosocial values were central to how she viewed the profession:
    \begin{quote}``We can tell them here is why you're doing this. You can have an impact in the world - not just making a salary [...] but bringing a benefit to society, which is what all engineering is all about, right?"
    \end{quote}
    \item \textbf{Personal Interest}: Several instructors mentioned personal interest as a motivation for talking about sustainability. For Andrew for example, this process happened more organically: ``it comes up just because it's something that kind of is in my mind".
    \item \textbf{Addressing Knowledge Gaps:} Several instructors described noticing gaps in the students' knowledge, or sustainability competencies, and increasing their coverage of sustainability in response. For example, Andrew's motivation ``comes from a number of years of reviewing work by other engineers and thinking: why aren't you thinking about these things?".
\end{itemize}

\subsubsection{Student Reception}\label{sect:results_reception}
Instructors described two distinct experiences when it came to student reception. 
Several instructors, particularly those integrating discussions of sustainability into technical coursework, described largely engaged students, who cared about the climate crisis and wanted to contribute to climate action: \begin{itemize}
    \item  Andrew: ``The students are generally quite receptive to it. They're interested in hearing some of those things."
    \item Amir: ``This is something that actually most people love hearing about because most of our students
are actually quite aware of the trajectory of climate change and sustainable development dangers
that are ahead of us."
    \item Maria: ``[students] were, like, thirsty for this lens and how they could contribute to sustainability."
\end{itemize}

Other instructors described mixed reactions, with some students engaged and others more resistant, perceiving such courses as an `easy A', a box-ticking requirement or not in line with their perceptions of engineering identity: 
\begin{itemize}
\item Chris: ``[Some students] already
have a predetermined idea of what is engineering or not. And if you try to get outside of that, then
we get into the preachy or activist or whatever label they attach to what we do."
\item Matt: ``It's on us really. The way sustainability is presented is an add-on. You take this course, you get your degree requirement, and then you go back to the technical stuff."
\item Arjun: ``[Certain courses] have a reputation of being easy and that's why many students try to
migrate towards these courses to just fulfill their requirements. I've heard feedback on that from some students who think that's a waste of time and
people shouldn't take these courses."
\end{itemize}

An interesting theme that emerged quite clearly in discussions of sustainability education related student experience in sustainability education to dominant ideas of disciplinary identity, and in particular, \textit{technicality}. Instructors expressed a range of perspectives. On one end of the spectrum, instructors were critical of less technical forms of sustainability education based on perceived substantivity:
\begin{quote}
\textbf{Arjun}: Some students seem to appreciate the types of courses we have introduced [...] that are a bit more substantive. And we don't allow students to take certain
other courses for that reason, because we don't think they're as valuable and students would
benefit much more from more in-depth education in that area. It stops them from getting more substantial, related education. Yeah, it's
a lost opportunity.
\newline
\textbf{Interviewer}: What do you think makes some courses more substantial than others in this context?
\newline
\textbf{Arjun}: I
mean, they have technical depth. So  I'd say they're less superficial compared to other
courses. But yeah, we have real technical knowledge that students learn compared to more
simple things like... I don't know what they're actually doing.
\end{quote}
Amir described a similar perspective on the value of technical, quantifiable sustainability education, describing how emphasis on empirical analysis and quantifiable data helps students to distinguish greenwashing from meaningful climate action:
\begin{quote}\textbf{Amir}: ``We're in the context of engineering. For engineering things, it's important to be
able to quantify, right? So we can't talk about abstract things. And when things become abstract,
us engineers, you know, we aren't good at it, right? But as soon as you tell people, you provide
them with numbers and you tell them methodology on how to get there or how to quantify their
impact, I think usually it flies very well in engineering classrooms."\end{quote}

On the other end of the spectrum, other instructors emphasize the importance of outcomes like leadership and agency. For example, Matt emphasizes agency as a key objective of his approach:
\begin{quote}
 \textbf{Matt}: I want to have them recognize their agency. You \textit{do} have expertise. You \textit{do} have an ability to add new ideas or question or things like that, because in five, ten years, hopefully many of these students will be in influential roles and be able to make decisions... I think a lot of students see engineers as problem solvers, not agents for change or having that sort of autonomy.
\end{quote}

Instructors also discussed how rigid student perceptions of engineering identity (viewing engineers as technical problem solvers versus social agents) could undermine their efforts to teach these types of outcomes: 

\begin{quote}
\textbf{Chris:} When we're talking about, you know, environmental regulations or process efficiency [...] This is practical, very
obviously linked to what students see as a role of engineers. So that's not a hard sell.
What becomes a harder sell is when you look at anything beyond that, anything more kind of
human, the social sustainability, right? And when you start talking about how human beings
connect to the engineering work [...], when we try to look at these things like social sustainability and the cultural factors and design and social factors and political factors and how all of this is part of the space in which we do engineering... There are students that look at this and say \textit{`this is not engineering. Why are we doing this? This is not anything that we're ever going to have to do'}. And so there's a level of disengagement.
\end{quote}

In Section \ref{sect:discuss_conceptualization}, we discuss how these instructor and student conceptualizations of engineering, and their perceptions of the `technicality' and `substantivity' of sustainability education relate to broader cultural patterns in engineering.  

\subsubsection{Barriers}
\label{sect:instr_barriers}
Several themes emerged in discussions around logistical barriers preventing more (or better) sustainability education in ECE:  
\begin{itemize}
    \item \textbf{Workload and Packed Curricula:} Engineering curricula are packed with content, and students already have an intense workload. Instructors described how in second and third year ECE courses in particular, chains of post-requisite courses depended on specific material being taught, making it difficult for instructors to change course content without disrupting tightly planned curricula. This was the primary reason Arjun did not discuss sustainability in his courses (``I'm afraid I wouldn't be able to cover all the other 
aspects of the course").
    \item \textbf{Teaching Evolving Fields and Topics} Instructors described sustainable computing as a quickly evolving area, which made these topics harder to include in a textbook or curriculum.
\item \textbf{Lack of Coordination}: Several instructors described a lack of coordination between faculty when it came to sustainability (``I honestly don't know who else is doing what"), making it difficult to develop a cohesive curriculum.
\end{itemize}

\section{Discussions}\label{sect:discussion}
We begin by discussing how instructors conceptualize sustainability in engineering (Section \ref{sect:discuss_conceptualization}), followed with an overview of the landscape of sustainability education in ECE (Section \ref{sect:discuss_landscape}). We then discuss opportunities for change (Section \ref{sect:discuss_challenges_opportunities}), and conclude with a discussion of limitations and threats to validity (Section \ref{sect:discuss_validity}).

\subsection{RQ1: Conceptualizations of Sustainability Education} \label{sect:discuss_conceptualization}
The premise of our work is that addressing the social and environmental issues associated with computer and electronics industries requires for universities to improve sustainability education in ECE programs. Many instructors and students will agree with this high level argument, but as we consider more concretely what ``improving sustainability education'' entails it is clear that the phrase is slippery, and value-laden: What exactly constitutes `sustainability education'? What does `quality' mean in this context? Who is in a position to make these evaluations and judgments? Decisions around which competencies are legitimate in engineering education are entangled with cultural and institutional values and norms. 

One instructor's use of the word ``substantive" to argue that sustainability eduction ought to involve technical problem solving and quantitative analysis is revealing. The term substantive (according to the Cambridge Dictionary, ``important, serious, or related to real facts")  links importance with positivism and objectivity. This is a common thread across several interviews, which argue that improving sustainability education means making it more `technical', `quantitative' and `practical'. Positivism is, as many researchers \citep{cech2015depoliticization, erden2003positivistic, niles2018bringing, riley2022engineering} (and our interviewees) explain, the dominant epistemological perspective in engineering departments, where students learn to privilege technocentric and objective knowledge, while their public welfare concerns have been shown to decline \citep{cech2014culture}. In our analysis of program descriptions, this technocentrism is particularly evident for ECE programs, whose program descriptions for prospective students are unlikely to discuss sustainability or include the pro-social calls to action that are more common in other disciplines. It is unsurprising then, that the educators interviewed who taught more critical perspectives or used teaching strategies that involved more open-ended subjective discussions, described resistance from students who ``look at this and say: this is
not engineering". 

In engineering, and in ECE in particular (which, as we show, has the least coverage of social and environmental sustainability), positivism and techno-solutionism are the water we swim in. Engineering educators who teach more sociotechnical, critical or subjective understandings are swimming against this current, facing barriers to student engagement, critical reception from other educators, and conflicting messaging from the `hidden curriculum' of cultural norms. Educators who focus on integrating sustainability through technical quantitative problems and activities are swimming with this current, and face less student resistance.

Many of the problems students are most likely to face however, are not only technical but also political and  subjective. They will need to navigate a technology industry associated with increasing social and environmental harm. Whether they understand their own choices in this context as political decisions, personal decisions, or engineering decisions, challenging extractive industries and practices requires not only technical skills but courage, solidarity and critical hope. In their work on critical engineering education, Claris and Riley (\citeyear{claris2012situation}) write that ``engineers may have abundant logical thinking skills but no practice of thinking critically about problem framing, power relations within the profession, hegemonic
epistemologies of the discipline, or reproductive practices of engineering education.
This leaves society with technologies that replicate power relations of the status
quo." Addressing sustainability requires disrupting this status quo, and therefore means training students to question design paradigms that reinforce harmful systems and patterns, and to engage in concrete personal action.

Pragmatically, we argue for both more technical and more critical sustainability education in ECE. However, for instructors and students to be able to draw connections between technical sustainable development skills and critical understandings, epistemic pluralism and cooperation is required. As Cech and Sherick (\citeyear{cech2015depoliticization}) point out in their work on sociotechnical engineering education, siloing of sociopolitical and technical knowledge reinforces a sense of technical/social dualism and leaves students ``unprepared to deal with
socio-cultural complexities inherent in real world engineering [and lacking] the intellectual tools and epistemological scaffolding necessary to clearly recognize such complexities". In preventing students and educators from effectively sharing their perspectives and teaching philosophies, a lack of discussion and encouragement of a broader range of perspectives within ECE departments forms a barrier to more coherent sustainability education. 

\subsection{RQ2: The Existing Canadian Engineering Education Landscape} \label{sect:discuss_landscape}
A key finding from our analysis of program documents is the marked difference between disciplines when it comes to integrating sustainability, despite having identical accreditation requirements. Mechanical, Electrical and Computer Engineering programs primarily (often exclusively) introduce sustainability through extra-departmental courses, while in Chemical, Civil and Environmental Engineering departments, many technical electives typically also discuss sustainable development. This difference means ECE students are receiving less sustainability education, and are learning different sustainability competencies: shared introductory courses primarily cover design, systems thinking and sustainability literacy, as opposed to more concrete and technical skills-based competencies. 
 Our analysis of program descriptions also reveals a dramatic difference in how different programs present the role of the engineer. In Chemical, Civil and Environmental Engineering, sustainability is more likely to be presented as a core component of engineering identity. In Mechanical, Electrical and Computer Engineering, sustainability (if mentioned at all) is presented as one of many possible applications of skills or theories, or as one of many design metrics. 

 ECE students are likely to only hear about sustainability in courses offered by other engineering departments or faculties. It is unsurprising then if they see sustainability as disconnected from their work, and difficult to practically apply. Students ostensibly learn in extra-departmental courses on professionalism or socio-environmental impact, that sustainability is a central professional responsibility, but are not shown the connection between that responsibility and their technical work. As Cech and Sherick (\citeyear{cech2015depoliticization}) write in their work on depoliticization in engineering education, ``by peeling off ethics training into a separate
course and codifying the relative unimportance of professional/ethical responsibilities by requiring students to take only one course on the topic, this arrangement
likely reinforces, rather than undermines, the ideology of depoliticization". A similar logic applies to sustainability education in ECE: a lack of discussion of socio-environmental dimensions within ECE  reinforces the perception that these problems are not the concern of electrical or computer engineers, which in turn undermines existing efforts at sustainability education. 

\subsection{RQ3: Barriers, Opportunities and Change Strategies} 
\label{sect:discuss_challenges_opportunities}
We summarize our findings by identifying six opportunities to change the landscape of ECE sustainability education, alongside the `change strategy' \citep{borrego2014increasing} they represent.   

\subsubsection{Integrate sustainability into existing technical ECE courses (Diffusion)} \label{sect:discuss_changes1}
In Borrego and Henderson's  (\citeyear{borrego2014increasing}) framework of change strategies, diffusion refers to broad adaptation and adoption of existing pedagogical interventions and innovations. In the context of sustainability education in ECE, we note that the substantial differences in sustainability education between disciplines (Section \ref{sect:discuss_landscape}) provides an opportunity for ECE educators to adopt existing strategies from different disciplines. In Table \ref{tab:technical_integration}, we include a list of opportunities for integration of sustainability throughout the ECE curriculum, by adopting ideas from existing ECE courses, and adapting ideas from non-ECE courses. %Publication of open education resources related to sustainability education in engineering could help facilitate this sort of diffusion. 

These additions do not need to be major, or formalized in the syllabus or course description. As one instructor described, rather than adding new modules or topics, sustainability can also be integrated a more granular level throughout a course, by taking discussions of performance or design decisions and ``just stretching the conversation a little bit [...] to think of the broader system and this whole ecosystem of different factors that affect decision making". 

%TC:ignore
\begin{table}[htb!]
\caption{Possibilities for integration of sustainability into technical ECE courses, based on inspiration from existing ECE courses, non-ECE courses and instructor interviews}
\label{tab:technical_integration}
\centering
\fontsize{8pt}{10pt}\selectfont
  \begin{tabular}{p{1.4cm}| p{7.9cm} p{5.3cm}}
  Course & Inspiration (either \color{red}{direct ECE examples}\color{black}, \color{orange}{interviewee ideas} \color{black} or \color{blue}non-ECE parallels\color{black}) & Curricular Change \\ \hline
Power Systems & ``First, various means of electric power generation-through hydroelectric, thermoelectric, geothermal, wind, solar, and nuclear sources are highlighted, and \textbf{the choice of a given source-dictated by economic and environmental factors} [...]" \textit{\color{red}(ELEE3260U, OTU)} & In the context of power generation, discuss evaluation of quantifiable environmental tradeoffs, or include broader discussions of social and environmental context.  \\ \hline
Energy \mbox{Conversion} / Power \mbox{Electronics} & ``Motors and generators are used in daily life for energy conversion, transportation, and within industry, and hence are an important part of the field of electrical engineering. This course explains the operation of electric machines, \textbf{including machines used in electric vehicles and renewable energy systems} [...]"  \textit{\color{red}(EEE231, RMC)} & In courses on energy conversion or power electronics, cover technologies used in renewable energy systems, highlight applications relevant to renewable energy systems. \\ \hline
Smart Grid & ``The course will focus on the role/impacts of the various smart grid components on the electric energy systems,  \textbf{including renewable, plug-in hybrid electric vehicles, demand side management, and greenhouse gas (GHG) emissions reductions}." \textit{\color{red}(ELEE4115U, UofT)} & In discussions of the smart grid, or smart systems, highlight and quantify environmental benefits like GHG emission reductions (and flag potential environmental concerns or tradeoffs). \\ \hline
Material \mbox{Selection} & ``Materials for engineering, atomic bondings, crystalline structures, properties of metals, glasses, semiconductors, insulators and magnetic materials. Electronic conduction in solids and simple devices. Materials in engineering design  \textbf{and environmental effects}." \textit{\color{red}(ECE220, UVic)} & Outline and evaluate the environmental and social considerations associated with material selection. Discuss the embodied carbon of materials, or the social and environmental impacts of supply chains. \\ \hline 
%Circuit \mbox{Analysis} & \textit{\fontsize{10pt}{12pt}\selectfont ``In second year when we do circuits... Well \textbf{modeling solar panels or solar cells} is very much just using basic circuits, diode, resistors - everything that you would learn in a first course in circuits. That could be a small unit or even like part of a lecture." \color{orange}\textit{(Interview with `Maria')}}  & \fontsize{10pt}{10pt}\selectfont Highlight applications relevant to sustainable development, even in early courses on circuit design and analysis. \\ \hline
Introduction to VLSI & \textit{``Students are quite familiar with the notion of measuring performance, energy and power. But things like carbon
emissions, it's a bit nuanced, right? So \textbf{I have to get into discussion of embodied versus operational and explain the lifecycle of the device}." \color{orange}\textit{(Interview with `Arjun')}}& In courses on circuits or computer architecture, discuss the product lifecycle, including impacts of ewaste, mining, operational carbon and embodied carbon. Introduce lifecycle analysis. \\ \hline
Computer \mbox{Architecture} & ``This course deals with sustainable product design [...] \textbf{Design topics include: product design for manufacture and assembly, design for disassembly and design for environment. Product end-of-life considerations include: recycling, remanufacture and reuse.} [...]" \color{blue}\textit{(MECH424, Queen's)} & In courses on computer architecture (or design of any consumer electronics), discuss end-of-life considerations. Outline how products can be made more durable and repairable, and how these design decisions relate to sustainability. \\ \hline
High \mbox{Performance} Computing, Datacenter Design & ``Impacts of architecture on the environment; \textbf{ecologic footprint, energy consumption, air quality, waste generation; designing with the environment; renewable energy,} effective siting and landscape, passive solar energy, natural lighting, energy efficiency." \textit{\color{blue}(ENVE1001, Carleton)} & Discuss the environmental impacts of large scale computing projects (distributed or in datacenters), including operational emissions, embodied emissions, energy sources and impacts associated with cooling.\\ \hline
Software \mbox{Engineering} & ``This course introduces fundamental principles of ecological engineering and their \textbf{application to understand and assess issues related to the provision of basic urban services through nature-based solutions}. Topics include the role of ecosystem services in urban processes; socio-ecological systems; sustainable and low-impact urban development; resilience-building capacities; [...]" \color{blue}\textit{(ENG421, UNBC)} & In software engineering courses, discuss applications of software engineering that facilitate climate action and sustainable development. Discuss case studies that show how software can have a positive social or environmental impact.\\ \hline
Human Computer Interaction & ``[...] Sustainability-minded products should be technically efficient, but also support people to use them more sustainably. Finally, many products and systems nudge people to behave in ways contrary to the user's best interests. \textbf{To address the above, the course focuses on design that increases intended product use, and pro-social / pro-environmental behaviours.}" \color{blue}\textit{(MIE440, UofT)} & In discussions of HCI, highlight how computing systems can be designed in a way that promotes sustainable use and pro-social behaviours - or unsustainable overuse. \\ \hline
Artificial \mbox{Intelligence} & Mechanical design of vehicular engines for different applications. Gas exchange and combustion engine processes. Combustion chambers design. [...] Engine testing. \textbf{Environmental impact of vehicular engines on global pollution. Recent developments in energy efficient and “clean” engines.} \color{blue}\textit{(MECH454, Concordia)}.& 
Discuss not only the algorithms and theory behind machine learning systems, but also the impact of these technologies on society and the environment. Compare the energy used during training or inference, between different systems, architectures or hardware platforms.
  \end{tabular}
\end{table}
%TC:endignore
\subsubsection{Create ECE-specific courses on social and environmental impact and professional practice (Organizational Development, Complexity Leadership)} 

The perceived lack of relevance of extra-departmental courses dedicated to sustainability (discussed in Section \ref{sect:discuss_landscape}) could be addressed by substituting (or supplementing) general engineering `impacts' courses with ECE-specific equivalents that more directly relate technical ECE content to sustainable development. For example, instead of taking a general introduction to engineering professional practice, ECE190 at Waterloo introduces engineering professionalism and sustainability in the context of ECE:
\begin{quote}
``\textbf{Engineering Profession and Practice}: Introduction to electrical and computer engineering with an emphasis on the profession of engineering, and engineering design. Topics include engineering design, safety, risk analysis, engineering data analysis, project management, \textit{\textbf{sustainability}}, business, entrepreneurship, and intellectual property. [...]"
\end{quote}

This approach is most in line with Borrego and Henderson's (\citeyear{borrego2014increasing}) `Organizational Development' strategy, in that it requires for departmental leadership to make major changes to the curriculum to bring it into alignment with a new vision.

\subsubsection{Reevaluate ECE recruitment materials (Organizational Development, Complexity Leadership)} \label{sect:discuss_changes3}
These materials impact which students choose ECE, and therefore determine the interests and motivations of incoming students. As discussed in Section \ref{sect:discuss_landscape}, in ECE, such materials include substantially less discussion of sustainability than other disciplines. As one instructor pointed out, without more explicit and central discussions of sustainability, ``the sector of students who care about the climate, these folks are going to go elsewhere."  Such public facing program materials are also part of a hidden curriculum of lessons about engineering identity that are communicated to students throughout their degrees. The absence of discussions of sustainability or social good in typical ECE program descriptions implicitly communicates that these are not central concerns in ECE, and presents engineers as technocrats rather than social (or  sociotechnical) agents. 

This approach could be considered `Organizational Development', in that it requires for departmental leadership to make changes to program materials and promotion to align them with a different vision for the program. Given that these changes attempt to stimulate cultural change in a complex system by disrupting existing patterns, they could also be said to be an example of `Complexity Leadership' \citep{borrego2014increasing}.

\subsubsection{Strengthen sustainability-related accreditation criteria (Quality Assurance)}  \label{sect:discuss_change strat}
In Section \ref{sect:accred_analysis}, we discuss how internationally, several accreditation systems include references to sustainability in Graduate Attributes related to ethics and problem analysis. Engineers Canada could follow this lead and integrate discussions of sustainability through more of its Graduate Attributes. This is not just a matter of reinforcing the importance of sustainability through more references, but also of including a wider range of sustainable development competencies, and perspectives on sustainable development.

Prior research argues that anti-colonial education supports sustainability by presenting alternatives to extractive colonial views and practices \citep{jimenez2023indigenizing, druker2022integration}, and 
Engineers Canada itself calls for ``transforming the engineering academic institution into one in which Indigenous epistemologies, ontologies, and cultures are meaningfully woven into the fabric of the engineering academic system" \citep{engcanreport2022}. Accreditation criteria do not acknowledge decolonization however, and we did not see Indigenous perspectives discussed in any Canadian ECE course descriptions. We see an opportunity here to explore how decolonizing accreditation could support decolonizing engineering education, and in turn, sustainable development. 

Accreditation-based approaches to integrating sustainability into curricula are clear examples of Borrego and Henderson's (\citeyear{borrego2014increasing}) `Quality Assurance' change strategy.

\subsubsection{Develop sustainability-oriented communities of practice (Faculty Learning Communities)} 
Instructors describe lack of coordination between faculty as a barrier to developing a cohesive sustainability curriculum, particularly given packed and interdependent engineering curricula (Section \ref{sect:instr_barriers}). Creating teaching communities within ECE focused on sustainable development could help educators share ideas and coordinate coherent sustainability education across the curriculum. 

Fostering a collaborative community of practice is an example of Borrego and Henderson's (\citeyear{borrego2014increasing}) `Faculty Learning Communities' strategy. 

\subsubsection{Bring transformative and critical pedagogy into ECE (Complexity Leadership)} \label{sect:recc_cp}
Meaningfully reorganizing engineering practice around social and ecological responsibility requires transformative sustainability education: if transformative education is the goal, then it is not sufficient to make incremental changes to course content without questioning fundamentally unsustainable engineering paradigms. Explicit mention of critical pedagogies or inclusion of a broader range of perspectives (including feminist perspectives, post-colonial approaches, post-growth philosophies, indigenous knowledge systems) was conspicuously absent from ECE programs in particular. Specific examples of programs and courses that subvert dominant approaches  (eg: “Two-Eyed Seeing [...] will be used to study the evolution of Indigenous design and technology and in the development of new sustainable technologies.” - ENGR4570, Ontario Technical University) are exceptions that prove the rule. We emphasize that any absence of discussion of these approaches to sustainability pedagogy in this paper does not reflect their lack of importance or relevance, but rather their scarcity.   

Borrego and Henderson present `Complexity Leadership' as a strategy for fostering emergence of novel ideas by disrupting of existing patterns \citep{borrego2014increasing}. Certainly integration of a broader range of critical pedagogies and perspectives into ECE would be a disruption, and we hope to see future work continue to explore the outcomes of such approaches.   

\subsection{Research Validity and Limitations}\label{sect:discuss_validity}

Threats to data validity are discussed throughout this paper, but we note in particular a few sources of potential bias or skewed data, and how they were mitigated:
\paragraph{Participant Selection:} 
    We contacted educators through our own personal networks, rather than through a more systematic process, although we did attempt to contact a range of different instructors with different levels of experience. We believe our triangulation through mixed methods helps to mitigate this limitation, by integrating more systematic analysis across Canadian universities and teaching communities. 
\paragraph{Researcher Bias and Subjectivity:} Our opinions on the importance of sustainability education is necessary in ECE, and our teaching philosophies color our discussions throughout this paper, and could bleed into interviews and data analysis. 
    We took concrete measures to mitigate issues of validity related to personal bias, and our dual status as researchers, educators and students by:
    \begin{enumerate} 
    \item Building our awareness of diverse pedagogical perspectives through cross-disciplinary readings and collaborations.
    \item Working reflexively, continuously writing about, and discussing our own personal positions and thoughts on education, and how those positions related to this research.
    \item Communicating our own positionalities (\Cref{sect:positionality}).
    \end{enumerate}
   % \item \textbf{Fair Dealing and Attention to Negative Cases}: To incorporate a wide range of perspectives, we interviewed instructors with a range of opinions and experiences teaching sustainability. Throughout the findings and discussions, we discuss how different participants had different perspectives and experiences. We pay close attention to conflicting elements of data in Sections \ref{sect:discuss_conceptualization}, \ref{sect:results_reception} and \ref{sect:recc_cp} in particular.
\paragraph{Issues of External Reliability, Limitation of Scope:} Another limitation of this study is that the scope of data collection was limited to English-language Canadian engineering programs. Due to language barriers, we excluded French-Canadian (and international) programs. Many of our results may be specific to our Canadian context, and are certainly not universal. %, relating closely to the broader cultural, political and economic context. 
    
\paragraph{Issues of Internal Reliability} Individual data sources like course descriptions may not always be fully descriptive, accurate, or correctly interpreted, and do not necessarily reflect actual student learning. We mitigate this issue by supplementing our analysis of course descriptions with instructor interviews, and by reporting inter-rater reliability.

\section{Conclusions and Future Work}\label{sect:conclusion}
Our research traces the landscape of sustainability education in ECE departments, describing the existing condition through quantitative and qualitative lenses, and identifying the mechanisms that shape it. We perform a large scale analysis of program, accreditation and course materials, which we use not only to evaluate content and curriculum structures, but also as cultural artifacts that reveal what students learn about disciplinary identity both explicitly and implicitly throughout their degrees. Tracing how this landscape changes across engineering disciplines and accreditation jurisdictions reveals how the existing curricula are shaped by distinct cultural norms and policies. Our interviews with instructors %and student survey 
contextualize these curricular structures, explaining how they relate to the lived experiences and perspectives of the people who make up ECE. 

Definitions of `substantive sustainability education' are subjective and negotiated across different systems of knowledge. Combining quantitative and qualitative analysis therefore allows us to not only describe this landscape, but to discuss how it is shaped both intentionally and through implicit cultural norms.  
In line with our mixed research methodology, we call for changing sustainability education in ECE not only through concrete curriculum changes, but through changes to the hidden curriculum. ECE program descriptions that seldom discuss sustainability may fail to reach some prospective students who hope to address social and environmental problems through engineering. Making these connections visible, and providing corresponding opportunities in the curriculum, could invite students with a wider range of motivations to consider ECE. Our study does not establish whether these messages influence applications or cohort composition; future research could examine those outcomes and students’ experiences after enrollment.

ECE needs more sustainability coverage, and more initiatives that disrupt binaries of technical/social, engineering/non-engineering which form barriers to cooperation between instructors and across departments. In our ongoing work on critical pedagogy in ECE, on electronic repair as form of sustainability education, and on integrating lightweight interventions into the ECE curriculum, we will continue to explore the opportunities described in Section \ref{sect:discuss_challenges_opportunities}. 

We conclude with a call to action for instructors, students, education researchers, department leaders and curriculum committees: addressing the climate crisis and confronting the social and environmental impacts associated with computing is an all-hands-on-deck situation. Particularly as Generative AI becomes increasingly prevalent,  teaching ECE students to understand the implications of these technologies, and their work, is critical. Among the engineering disciplines studied, ECE has the least coverage of sustainability, and its program materials implicitly tell students that sustainability is not a central concern of the field. Closing these gaps is urgent, and requires coordinated effort from all groups.

\section{Positionality}\label{sect:positionality}
Our interest in this work stems from our previous work teaching ECE students about sustainability, and our own experiences navigating sustainable development as Electrical/Computer engineers. We work from within an ECE department at a large Canadian university, and our educational and professional backgrounds are largely technical. We therefore analyze Canadian ECE education through an insider's lens, and our perspectives are shaped in part by the traditionally analytical and technical perspectives that are dominate ECE. Our teaching philosophies and perspectives on engineering education are also shaped by our interest in more critical pedagogies however, and by our collaborations with interdisciplinary and sustainability-focused communities of practice. We note that the first author, who conducted all interviews, is a PhD student, and in no position of power or authority over the educators interviewed.

%Esther is a PhD candidate in an ECE department. Her interest in sustainability education grew, in part, from her own experiences struggling to reconcile complex socio-technical problems with a highly technical engineering background. Esther is a part of the ECE graduate student community, and given her background and program, has an insider's lens into ECE education. However, given her interdisciplinary PhD research on sustainable computing and engineering education, and the fact that she is neither faculty nor an undergraduate student, she also holds an outsider's position with respect to the interviewees and survey participants. %Esther's perspectives on pedagogy lean towards critical pedagogy, but are informed by her own technical education, and her pragmatic perspective is likely also informed by an engineering education that underscored practicality.  
%Sathish...

\section{Statement on Artificial Intelligence}\label{sect:ai}
Artificial intelligence (claude-sonnet-4.6) was used to categorize course descriptions by sustainability competency (as discussed in Section \ref{sect:doc_analysis}). 
Artificial intelligence was not used to analyze any other data or to generate written content. 

\bibliographystyle{apalike}
\bibliography{bibliography}

@article{rockstrom2009planetary,
  title={Planetary boundaries: exploring the safe operating space for humanity},
  author={Rockstr{\"o}m, Johan and Steffen, Will and Noone, Kevin and Persson, {\AA}sa and Chapin III, F Stuart and Lambin, Eric and Lenton, Timothy M and Scheffer, Marten and Folke, Carl and Schellnhuber, Hans Joachim and others},
  journal={Ecology and society},
  volume={14},
  number={2},
  year={2009},
  publisher={JSTOR}
}

@incollection{cech2015depoliticization,
  title={Depoliticization and the structure of engineering education},
  author={Cech, Erin A and Sherick, Heidi M},
  booktitle={International Perspectives on Engineering Education: Engineering Education and Practice in Context, Volume 1},
  pages={203--216},
  year={2015},
  publisher={Springer}
}

@article{klinger2018rare,
  title={Rare earth elements: Development, sustainability and policy issues},
  author={Klinger, Julie Michelle},
  journal={The Extractive Industries and Society},
  volume={5},
  number={1},
  pages={1--7},
  year={2018},
  publisher={Elsevier}
}

@article{grant2013health,
  title={Health consequences of exposure to e-waste: a systematic review},
  author={Grant, Kristen and Goldizen, Fiona C and Sly, Peter D and Brune, Marie-Noel and Neira, Maria and van den Berg, Martin and Norman, Rosana E},
  journal={The lancet global health},
  volume={1},
  number={6},
  pages={e350--e361},
  year={2013},
  publisher={Elsevier}
}

@misc{balde2024global,
  title={Global e-waste monitor 2024},
  author={Bald{\'e}, Cornelis P and Kuehr, Ruediger and Yamamoto, Tales and McDonald, Rosie and D’Angelo, Elena and Althaf, Shahana and Bel, Garam and Deubzer, Otmar and Fernandez-Cubillo, Elena and Forti, Vanessa and others},
  year={2024},
  publisher={International Telecommunication Union (ITU) and United Nations Institute for~…}
}

@article{kollar2026fossil,
  title={The fossil-AI nexus: petrostate capitalism, computing power, and the production of powered land},
  author={Kollar, Justin},
  journal={New Political Economy},
  pages={1--22},
  year={2026},
  publisher={Taylor \& Francis}
}

@article{peters2024sustainability,
  title={Sustainability in computing education: A systematic literature review},
  author={Peters, Anne-Kathrin and Capilla, Rafael and Coroam{\u{a}}, Vlad Constantin and Heldal, Rogardt and Lago, Patricia and Leifler, Ola and Moreira, Ana and Fernandes, Jo{\~a}o Paulo and Penzenstadler, Birgit and Porras, Jari and others},
  journal={ACM transactions on computing education},
  volume={24},
  number={1},
  pages={1--53},
  year={2024},
  publisher={ACM New York, NY}
}

@article{thurer2018systematic,
  title={A systematic review of the literature on integrating sustainability into engineering curricula},
  author={Th{\"u}rer, Matthias and Toma{\v{s}}evi{\'c}, Ivan and Stevenson, Mark and Qu, Ting and Huisingh, Don},
  journal={Journal of Cleaner Production},
  volume={181},
  pages={608--617},
  year={2018},
  publisher={Elsevier}
}

@article{weisser2017defining,
  title={Defining sustainability in higher education: a rhetorical analysis},
  author={Weisser, Christian R},
  journal={International Journal of Sustainability in Higher Education},
  volume={18},
  number={7},
  pages={1076--1089},
  year={2017},
  publisher={Emerald Publishing Limited}
}

@article{sundman2025experiential,
  title={Experiential learning for sustainability: A systematic review and research agenda for engineering education},
  author={Sundman, Julia and Feng, Xiaoqi and Shrestha, Anup and Johri, Aditya and Varis, Olli and Taka, Maija},
  journal={European Journal of Engineering Education},
  pages={1--31},
  year={2025},
  publisher={Taylor \& Francis}
}

@article{kahn2008education,
  title={From education for sustainable development to ecopedagogy: Sustaining capitalism or sustaining life},
  author={Kahn, Richard},
  journal={Green theory \& praxis: The journal of ecopedagogy},
  volume={4},
  number={1},
  year={2008}
}

@article{lozano2017connecting,
  title={Connecting competences and pedagogical approaches for sustainable development in higher education: A literature review and framework proposal},
  author={Lozano, Rodrigo and Merrill, Michelle Y and Sammalisto, Kaisu and Ceulemans, Kim and Lozano, Francisco J},
  journal={Sustainability},
  volume={9},
  number={10},
  pages={1889},
  year={2017},
  publisher={MDPI}
}

@article{ozkan2026making,
  title={Making sociotechnical doable: Cross-disciplinary translations into engineering education},
  author={{\"O}zkan, Desen S and Zacharias, Kari and Kleine, Marie Stettler},
  journal={Journal of Engineering Education},
  volume={115},
  number={4},
  pages={e70076},
  year={2026},
  publisher={Wiley Online Library}
}

@article{jimenez2023indigenizing,
  title={Indigenizing environmental sustainability curriculum and pedagogy: Confronting our global ecological crisis via Indigenous sustainabilities},
  author={Jimenez, Jeremy and Kabachnik, Peter},
  journal={Teaching in Higher Education},
  volume={28},
  number={5},
  pages={1095--1107},
  year={2023},
  publisher={Taylor \& Francis}
}

@article{rodriguez2018roadmapping,
  title={Roadmapping towards sustainability proficiency in engineering education},
  author={Rodriguez-Andara, Alejandro and R{\'\i}o-Belver, Rosa Maria and Rodr{\'\i}guez-Salvador, Marisela and Lezama-Nicol{\'a}s, Ren{\'e}},
  journal={International Journal of Sustainability in Higher Education},
  volume={19},
  number={2},
  pages={413--438},
  year={2018},
  publisher={Emerald Publishing Limited}
}

@article{alsubaie2015hidden,
  title={Hidden curriculum as one of current issue of curriculum.},
  author={Alsubaie, Merfat Ayesh},
  journal={Journal of Education and practice},
  volume={6},
  number={33},
  pages={125--128},
  year={2015},
  publisher={ERIC}
}

@article{henderson2011facilitating,
  title={Facilitating change in undergraduate STEM instructional practices: An analytic review of the literature},
  author={Henderson, Charles and Beach, Andrea and Finkelstein, Noah},
  journal={Journal of research in science teaching},
  volume={48},
  number={8},
  pages={952--984},
  year={2011},
  publisher={Wiley Online Library}
}

@article{ralph2020empirical,
  title={Empirical standards for software engineering research},
  author={Ralph, Paul and Ali, Nauman bin and Baltes, Sebastian and Bianculli, Domenico and Diaz, Jessica and Dittrich, Yvonne and Ernst, Neil and Felderer, Michael and Feldt, Robert and Filieri, Antonio and others},
  journal={arXiv preprint arXiv:2010.03525},
  year={2020}
}

@article{mays2000assessing,
  title={Assessing quality in qualitative research},
  author={Mays, Nicholas and Pope, Catherine},
  journal={Bmj},
  volume={320},
  number={7226},
  pages={50--52},
  year={2000},
  publisher={British Medical Journal Publishing Group}
}

@book{gillespie2024pragmatism,
  title={Pragmatism and methodology: Doing research that matters with mixed methods},
  author={Gillespie, Alex and Gl{\u{a}}veanu, Vlad and de Saint Laurent, Constance},
  year={2024},
  publisher={Cambridge University Press}
}

@article{klassen2020global,
  title={Do global norms matter? The new logics of engineering accreditation in Canadian universities},
  author={Klassen, Mike and S{\'a}, Creso},
  journal={Higher Education},
  volume={79},
  number={1},
  pages={159--174},
  year={2020},
  publisher={Springer}
}

@article{terry2017thematic,
  title={Thematic analysis},
  author={Terry, Gareth and Hayfield, Nikki and Clarke, Victoria and Braun, Virginia and others},
  journal={The SAGE handbook of qualitative research in psychology},
  volume={2},
  number={17-37},
  pages={25},
  year={2017},
  publisher={SAGE Publications Ltd}
}

@article{godfrey2010mapping,
  title={Mapping the cultural landscape in engineering education},
  author={Godfrey, Elizabeth and Parker, Lesley},
  journal={Journal of Engineering education},
  volume={99},
  number={1},
  pages={5--22},
  year={2010},
  publisher={Wiley Online Library}
}

@article{druker2022integration,
  title={Integration of indigenous and local knowledge into sustainability education: a systematic literature review},
  author={Druker-Ibanez, Sofia and Caceres-Jensen, Lizethly},
  journal={Environmental Education Research},
  volume={28},
  number={8},
  pages={1209--1236},
  year={2022},
  publisher={Taylor \& Francis}
}

@article{brundtland1987our,
  title={Our common future—Call for action},
  author={Brundtland, Gro Harlem},
  journal={Environmental conservation},
  volume={14},
  number={4},
  pages={291--294},
  year={1987},
  publisher={Cambridge University Press}
}

@techreport{ieaenergy,
  title={Energy and AI},
  author={{International Energy Agency}},
  institution={International Energy Agency},
  year={2025},
  address={Paris}
}

@article{mchugh2012interrater,
  title={Interrater reliability: the kappa statistic},
  author={McHugh, Mary L},
  journal={Biochemia medica},
  volume={22},
  number={3},
  pages={276--282},
  year={2012},
  publisher={Hrvatsko dru{\v{s}}tvo za medicinsku biokemiju i laboratorijsku medicinu}
}

@incollection{riley2022engineering,
  title={Engineering and social justice},
  author={Riley, Donna},
  booktitle={Engineering and social justice},
  pages={47--106},
  year={2022},
  publisher={Springer}
}

@article{cech2014culture,
  title={Culture of disengagement in engineering education?},
  author={Cech, Erin A},
  journal={Science, Technology, \& Human Values},
  volume={39},
  number={1},
  pages={42--72},
  year={2014},
  publisher={SAGE Publications, Sage CA: Los Angeles, CA}
}

@inproceedings{niles2018bringing,
  title={Bringing in" The Social": Resisting and Assisting Social Engagement in Engineering Education},
  author={Niles, Skye and Contreras, Santina and Roudbari, Shawhin and Kaminsky, Jessica and Harrison, Jill},
  booktitle={2018 World Engineering Education Forum-Global Engineering Deans Council (WEEF-GEDC)},
  pages={1--6},
  year={2018},
  organization={IEEE}
}

@article{ram2025crafting,
  title={Crafting a definition of sustainability for engineering education and applying it to assess curriculum},
  author={Ram, Sherry-Ann and Tihanyi, Deborah and MacLean, Heather L and Daniel Posen, I},
  journal={Sustainability Science and Technology},
  volume={2},
  number={2},
  pages={024004},
  year={2025},
  publisher={IOP Publishing}
}

@techreport{engcanreport2022,
  title={Report on truth and reconciliation in engineering education},
  author={{Engineers Canada}},
  institution={Engineers Canada},
  year={2022},
  address={Ottawa}
}

@techreport{gemproposed,
  title={Proposed gas-fired power plants in the United States rise due to AI energy demand speculation, but remain largely in early development stage},
  author={Martos, Jenny},
  year={2026},
  institution={Global Energy Monitor}
}

@article{gutierrez2021development,
  title={Development of contemporary engineering graduate attributes through open-ended problems and activities},
  author={Guti{\'e}rrez Ortiz, Francisco Javier and Fitzpatrick, John J and Byrne, Edmond P},
  journal={European Journal of Engineering Education},
  volume={46},
  number={3},
  pages={441--456},
  year={2021},
  publisher={Taylor \& Francis}
}

@article{wiek2011key,
  title={Key competencies in sustainability: a reference framework for academic program development},
  author={Wiek, Arnim and Withycombe, Lauren and Redman, Charles L},
  journal={Sustainability science},
  volume={6},
  number={2},
  pages={203--218},
  year={2011},
  publisher={Springer}
}

@article{brundiers2021key,
  title={Key competencies in sustainability in higher education—toward an agreed-upon reference framework},
  author={Brundiers, Katja and Barth, Matthias and Cebri{\'a}n, Gisela and Cohen, Matthew and Diaz, Liliana and Doucette-Remington, Sonya and Dripps, Weston and Habron, Geoffrey and Harr{\'e}, Niki and Jarchow, Meghann and others},
  journal={Sustainability science},
  volume={16},
  number={1},
  pages={13--29},
  year={2021},
  publisher={Springer}
}

@article{wang2022sustainability,
  title={Sustainability education at higher education institutions: pedagogies and students’ competences},
  author={Wang, Yijing and Sommier, M{\'e}lodine and Vasques, Ana},
  journal={International Journal of Sustainability in Higher Education},
  volume={23},
  number={8},
  pages={174--193},
  year={2022},
  publisher={Emerald Publishing Limited}
}

@article{gutierrez2022sustainability,
  title={Sustainability in engineering education: A review of learning outcomes},
  author={Gutierrez-Bucheli, Laura and Kidman, Gillian and Reid, Alan},
  journal={Journal of Cleaner Production},
  volume={330},
  pages={129734},
  year={2022},
  publisher={Elsevier}
}

@misc{eop,
  title={The engineering for one planet framework: Essential learning outcomes for engineering education},
  author={{Engineering One Planet}},
  year={2020},
  publisher={Engineering One Planet}
}

@inproceedings{godfrey2003theoretical,
  title={A theoretical model of the engineering education culture: A tool for change},
  author={Godfrey, Elizabeth},
  booktitle={2003 Annual Conference},
  pages={8--133},
  year={2003}
}

@article{gardelle2026sustainability,
  title={Sustainability in engineering education in Europe over the last 15 years: Visions and tendencies among educators as revealed by a literature review},
  author={Gardelle, Linda},
  journal={European Journal of Engineering Education},
  volume={51},
  number={4},
  pages={812--839},
  year={2026},
  publisher={Taylor \& Francis}
}

@article{borrego2014increasing,
  title={Increasing the use of evidence-based teaching in STEM higher education: A comparison of eight change strategies},
  author={Borrego, Maura and Henderson, Charles},
  journal={Journal of Engineering Education},
  volume={103},
  number={2},
  pages={220--252},
  year={2014},
  publisher={Wiley Online Library}
}

@article{prados2005quality,
  title={Quality assurance of engineering education through accreditation: The impact of Engineering Criteria 2000 and its global influence},
  author={Prados, John W and Peterson, George D and Lattuca, Lisa R},
  journal={Journal of engineering education},
  volume={94},
  number={1},
  pages={165--184},
  year={2005},
  publisher={Wiley Online Library}
}

@inproceedings{byrne2010engineering,
  title={Engineering education for sustainable development: a review of international progress},
  author={Byrne, Edmond P and Desha, Cheryl and Fitzpatrick, John J and Hargroves, Karlson},
  booktitle={International Symposium for Engineering Education},
  year={2010}
}

@article{byrne2023evolving,
  title={The evolving engineer; professional accreditation sustainability criteria and societal imperatives and norms},
  author={Byrne, Edmond P},
  journal={Education for Chemical Engineers},
  volume={43},
  pages={23--30},
  year={2023},
  publisher={Elsevier}
}

@article{byrne2013exploring,
  title={Exploring sustainability themes in engineering accreditation and curricula},
  author={Byrne, Edmond P and Desha, Cheryl J and Fitzpatrick, John J and “Charlie” Hargroves, Karlson},
  journal={International Journal of Sustainability in Higher Education},
  volume={14},
  number={4},
  pages={384--403},
  year={2013},
  publisher={Emerald Group Publishing Limited}
}

@incollection{stanivskis2014engineering,
  title={Engineering education for sustainable industries: Approach, implementation and accreditation},
  author={Stani{\v{s}}kis, Jurgis Kazimieras and Katili{\=u}t{\.e}, Egl{\.e}},
  booktitle={Transformative Approaches to Sustainable Development at Universities: Working Across Disciplines},
  pages={55--74},
  year={2014},
  publisher={Springer}
}

@incollection{freire2020pedagogy,
  title={Pedagogy of the oppressed},
  author={Freire, Paulo},
  booktitle={Toward a sociology of education},
  pages={374--386},
  year={1968},
  publisher={Routledge}
}

@article{claris2012situation,
  title={Situation critical: critical theory and critical thinking in engineering education},
  author={Claris, Lionel and Riley, Donna},
  journal={Engineering Studies},
  volume={4},
  number={2},
  pages={101--120},
  year={2012},
  publisher={Taylor \& Francis}
}

@article{mejia2018critical,
  title={Critical theoretical frameworks in engineering education: An anti-deficit and liberative approach},
  author={Mejia, Joel Alejandro and Revelo, Renata A and Villanueva, Idalis and Mejia, Janice},
  journal={Education Sciences},
  volume={8},
  number={4},
  pages={158},
  year={2018},
  publisher={MDPI}
}

@inproceedings{riley2012aiding,
  title={Aiding and ABETing: The bankruptcy of outcomes-based education as a change strategy},
  author={Riley, Donna M},
  booktitle={2012 ASEE Annual Conference \& Exposition},
  pages={25--141},
  year={2012}
}

@inproceedings{riley2003pedagogies,
  title={Pedagogies of liberation in an engineering thermodynamics class},
  author={Riley, Donna},
  booktitle={2003 Annual Conference},
  pages={8--918},
  year={2003}
}

@article{erden2003positivistic,
  title={Positivistic tendencies due to engineering education},
  author={Erden, Mustafa Suphi},
  journal={IFAC Proceedings Volumes},
  volume={36},
  number={10},
  pages={117--122},
  year={2003},
  publisher={Elsevier}
}

\appendix\label{app:appendix}
\section{Appendix}

%TC:ignore
\subsection{Program Descriptions}
\label{app:appendix_programs}
Our analysis of engineering program descriptions (Section \ref{sect:findings_program}), is based on public-facing program descriptions available on university websites. The URLs for each program descriptions are listed below. All were initially accessed and downloaded for analysis on November 19th, 2025.  

\begin{itemize}
    \item British Columbia Institute of Technology (BCIT)
    \begin{itemize}
    \item Civil Engineering: \url{https://www.bcit.ca/programs/civil-engineering-bachelor-of-engineering-full-time-8660beng/}
    \item Electrical Engineering: \url{https://www.bcit.ca/programs/electrical-engineering-bachelor-of-engineering-full-time-8030beng/}
    \item Mechanical Engineering: \url{https://www.bcit.ca/programs/mechanical-engineering-bachelor-of-engineering-full-time-8020beng/}
    \end{itemize}

    \item Carleton University (Carleton)
    \begin{itemize}
    \item Civil Engineering: \url{https://admissions.carleton.ca/programs/civil-engineering/}
\item Computer Systems: \url{https://admissions.carleton.ca/programs/computer-systems-engineering/}
\item Electrical Engineering: \url{https://admissions.carleton.ca/programs/electrical-engineering/}
\item Environmental: \url{https://admissions.carleton.ca/programs/environmental-engineering/}
\item Mechanical: \url{https://admissions.carleton.ca/programs/mechanical-engineering/}
    \end{itemize}
    
    \item Concordia University (Concordia)
    \begin{itemize}
    \item Civil Engineering:	\url{https://www.concordia.ca/academics/undergraduate/civil-engineering.html}
\item Computer Engineering: 	\url{https://www.concordia.ca/academics/undergraduate/computer-engineering.html}
\item Electrical Engineering: 	\url{https://www.concordia.ca/academics/undergraduate/electrical-engineering.html}
\item Mechanical Engineering: 	\url{https://www.concordia.ca/academics/undergraduate/mechanical-engineering.html}
    \end{itemize}
    
    \item Conestoga College Institute of Technology and Advanced Learning (Conestoga)
    \begin{itemize}
\item Electronic Systems Engineering: \url{https://www.conestogac.on.ca/fulltime/bachelor-of-engineering-electronic-systems-engineering}
\item Mechanical Systems Engineering: \url{https://www.conestogac.on.ca/fulltime/bachelor-of-engineering-mechanical-systems-engineering}
    \end{itemize}
    
    \item Dalhousie University (Dal)
    \begin{itemize}
    \item Chemical Engineering: 	\url{https://www.dal.ca/study/programs/undergraduate/chemical-engineering-beng.html}
\item Civil Engineering: 	\url{https://www.dal.ca/study/programs/undergraduate/civil-engineering-beng.html}
\item Electrical Engineering: 	\url{https://www.dal.ca/study/programs/undergraduate/electrical-engineering-beng.html}
\item Environmental Engineering:	\url{https://www.dal.ca/study/programs/undergraduate/environmental-engineering-beng.html}
\item Mechanical Engineering: 	\url{https://www.dal.ca/study/programs/undergraduate/mechanical-engineering-beng.html}
    \end{itemize}
    
    \item Lakehead University (LakeheadU)
    \begin{itemize}
    \item Chemical Engineering: 	\url{https://www.lakeheadu.ca/programs/departments/chemical-engineering/programs/ug}
\item Civil Engineering: 	\url{https://www.lakeheadu.ca/programs/departments/civil-engineering/programs/ug}
\item Electrical Engineering: 	\url{https://www.lakeheadu.ca/programs/departments/electrical-engineering}
\item Mechanical Engineering: 	\url{https://www.lakeheadu.ca/programs/departments/mechanical-engineering/programs/ug}
    \end{itemize}
    
    \item Laurentian University (Laurentian)
    \begin{itemize}
    \item Chemical Engineering: 	\url{https://laurentian.ca/academics/program/chemical-engineering}
\item Mechanical Engineering: 	\url{https://laurentian.ca/academics/program/mechanical-engineering}
    \end{itemize}
    
    \item McGill University (McGill)
    \begin{itemize}
    \item Chemical Engineering: 	\url{https://www.mcgill.ca/undergraduate-admissions/program/chemical-engineering-faculty-engineering}
\item Civil Engineering: 	\url{https://www.mcgill.ca/undergraduate-admissions/program/civil-engineering}
\item Computer Engineering: 	\url{https://www.mcgill.ca/undergraduate-admissions/program/computer-engineering}
\item Electrical Engineering: 	\url{https://www.mcgill.ca/undergraduate-admissions/program/electrical-engineering-faculty-engineering}
\item Mechanical Engineering: 	\url{https://www.mcgill.ca/undergraduate-admissions/program/mechanical-engineering}
    \end{itemize}
    
    \item McMaster University (McMaster)
    \begin{itemize}
    \item Chemical Engineering: 	\url{https://www.eng.mcmaster.ca/chemeng/degree-options/chemical-engineering/}
\item Civil Engineering: 	\url{https://www.eng.mcmaster.ca/civil/degree-options/civil-engineering/}
\item Computer Engineering: 	\url{https://www.eng.mcmaster.ca/ece/degree-options/computer-engineering/}
\item Electrical Engineering: 	\url{https://www.eng.mcmaster.ca/ece/degree-options/electrical-engineering/}
\item Mechanical Engineering: 	\url{https://www.eng.mcmaster.ca/mech/degree-options/mechanical-engineering/}
    \end{itemize}
    
    \item Memorial University of Newfoundland (MUN)
    \begin{itemize}
    \item Civil Engineering: 	\url{https://www.mun.ca/undergrad/programs/engineering/civil-engineering/}
\item Computer Engineering: 	\url{https://www.mun.ca/engineering/ece/future-students/}
\item Electrical Engineering: 	\url{https://www.mun.ca/engineering/ece/future-students/}
\item Mechanical Engineering: 	\url{https://www.mun.ca/engineering/mech/future-students/}
    \end{itemize}
    
    \item Queen's University (Queen's)
    \begin{itemize}
    \item Chemical Engineering: 	\url{https://smithengineering.queensu.ca/chee/undergraduate/chemical-engineering.html}
\item Civil Engineering: 	\url{https://smithengineering.queensu.ca/civil/undergraduate/civil-engineering.html}
\item Computer Engineering: 	\url{https://smithengineering.queensu.ca/ece/undergraduate/computer-engineering/index.html}
\item Electrical Engineering: 	\url{https://smithengineering.queensu.ca/ece/undergraduate/electrical-engineering/index.html}
\item Mechanical Engineering: 	\url{https://smithengineering.queensu.ca/mme/undergraduate/mechanical-materials-engineering.html}
    \end{itemize}
    
    \item Royal Military College of Canada (RMC)
    \begin{itemize}
    \item Chemical Engineering: 	\url{https://www.rmc-cmr.ca/en/registrars-office/undergraduate-chemistry-chemical-engineering-programmes}
\item Civil Engineering: 	\url{https://www.rmc-cmr.ca/en/registrars-office/undergraduate-civil-engineering-programme}
\item Computer Engineering: 	\url{https://www.rmc-cmr.ca/en/registrars-office/electrical-engineering-and-computer-engineering-undergraduate-programmes}
\item Electrical Engineering: 	\url{https://www.rmc-cmr.ca/en/registrars-office/electrical-engineering-and-computer-engineering-undergraduate-programmes}
\item Mechanical Engineering: 	\url{https://www.rmc-cmr.ca/en/registrars-office/mechanical-aeronautical-engineering-undergraduate-programmes}
    \end{itemize}
    
    \item Sheridan College Institute of Technology and Advanced Learning (Sheridan)
    \begin{itemize}
    \item Electrical Engineering: 	\url{https://www.sheridancollege.ca/programs/bachelor-of-engineering-electrical}
    \end{itemize}
    
    \item Toronto Metropolitan University (TMU)
    \begin{itemize}
\item Chemical Engineering:	\url{https://www.torontomu.ca/programs/undergraduate/chemical-engineering-co-op/}
\item Civil Engineering: 	\url{https://www.torontomu.ca/programs/undergraduate/civil-engineering/}
\item Computer Engineering: 	\url{https://www.torontomu.ca/programs/undergraduate/computer-engineering/}
\item Electrical Engineering: 	\url{https://www.torontomu.ca/programs/undergraduate/electrical-engineering/}
\item Mechanical Engineering: 	\url{https://www.torontomu.ca/programs/undergraduate/mechanical-engineering/}
    \end{itemize}
    
    \item University of Alberta (UofA)
    \begin{itemize}
    \item Chemical Engineering: 	\url{https://www.ualberta.ca/en/undergraduate-programs/bachelor-of-science-in-chemical-engineering-chemical-engineering.html}
\item Civil Engineering: 	\url{https://www.ualberta.ca/en/undergraduate-programs/bachelor-of-science-in-civil-engineering-civil-engineering.html}
\item Computer Engineering: 	\url{https://www.ualberta.ca/en/undergraduate-programs/bachelor-of-science-in-computer-engineering-computer-engineering.html}
\item Electrical Engineering: 	\url{https://www.ualberta.ca/en/undergraduate-programs/bachelor-of-science-in-electrical-engineering-electrical-engineering.html}
\item Mechanical Engineering: 	\url{https://www.ualberta.ca/en/undergraduate-programs/bachelor-of-science-in-mechanical-engineering-mechanical-engineering.html}
    \end{itemize}
    
    \item University of British Columbia (UBC)
    \begin{itemize}
    \item Chemical Engineering: \url{https://engineering.ubc.ca/programs/undergraduate/chemical-engineering}
\item Civil Engineering: 	\url{https://civil.ubc.ca/undergraduate/}
\item Computer Engineering: \url{https://ece.ubc.ca/undergraduates/programs/computer-engineering-program/}
\item Electrical Engineering: 	\url{https://ece.ubc.ca/undergraduates/programs/electrical-engineering-program/}
\item Environmental Engineering: \url{https://environmental.engineering.ubc.ca/}
\item Mechanical Engineering: 	\url{https://mech.ubc.ca/undergraduate/whymech/}
    \end{itemize}
    
    \item University of British Columbia - Okanagan (UBC-O)
    \begin{itemize}
    \item Civil Engineering: \url{https://engineering.ok.ubc.ca/programs-admissions/undergrad/civil/}
\item Electrical Engineering:  \url{https://engineering.ok.ubc.ca/programs-admissions/undergrad/electrical/}
\item Mechnical Engineering:	\url{https://engineering.ok.ubc.ca/programs-admissions/mechanical/}
    \end{itemize}
    
    \item University of Calgary (UCalgary)
    \begin{itemize}
    \item Chemical Engineering: 	\url{https://www.ucalgary.ca/future-students/undergraduate/explore-programs/chemical-engineering}
\item Civil Engineering: 	\url{https://www.ucalgary.ca/future-students/undergraduate/explore-programs/civil-engineering}
\item Electrical Engineering: 	\url{https://www.ucalgary.ca/future-students/undergraduate/explore-programs/electrical-engineering}
\item Mechanical Engineering: 	\url{https://www.ucalgary.ca/future-students/undergraduate/explore-programs/mechanical-engineering}
    \end{itemize}
    
    \item University of Guelph (UofG)
    \begin{itemize}
    \item Computer Engineering: 	\url{https://www.uoguelph.ca/programs/computer-engineering/}
\item Environmental Engineering: 	\url{https://www.uoguelph.ca/programs/environmental-engineering/}
\item Mechanical Engineering: 	\url{https://www.uoguelph.ca/programs/mechanical-engineering/}
    \end{itemize}
    
    \item University of Manitoba (UM)
    \begin{itemize}
    \item Civil Engineering: 	\url{https://umanitoba.ca/explore/programs-of-study/civil-engineering-bsc}
\item Computer Engineering: 	\url{https://umanitoba.ca/explore/programs-of-study/computer-engineering-bsc}
\item Electrical Engineering: 	\url{https://umanitoba.ca/explore/programs-of-study/electrical-engineering-bsc}
\item Mechanical Engineering: 	\url{https://umanitoba.ca/explore/programs-of-study/mechanical-engineering-bsc}
    \end{itemize}
    
    \item University of New Brunswick (UNB)
    \begin{itemize}
    \item Chemical Engineering: 	\url{https://www.unb.ca/fredericton/engineering/depts/chemical/undergrad.html}
\item Civil Engineering: \url{https://www.unb.ca/fredericton/engineering/depts/civil/undergrad.html}
\item Electrical Engineering: 	\url{https://www.unb.ca/fredericton/engineering/depts/ece/undergrad.html}
\item Mechanical Engineering: 	\url{https://www.unb.ca/fredericton/engineering/depts/mechanical/undergrad.html}
    \end{itemize}
    
    \item University of Northern British Columbia (UNBC)
    \begin{itemize}
    \item Civil Engineering: \url{https://www.unbc.ca/programs-and-admissions/undergraduate/programs/civil-engineering}
\item Environmental	Engineering: \url{https://www.unbc.ca/programs/environmental-engineering}
    \end{itemize}

    \item University of Ontario Institute of Technology (OntarioTech)
    \begin{itemize}
    \item Electrical Engineering: 	\url{https://ontariotechu.ca/programs/undergraduate/engineering/electrical-engineering/index.php}
\item Mechanical Engineering: 	\url{https://ontariotechu.ca/programs/undergraduate/engineering/mechanical-engineering/index.php}
    \end{itemize}
    
    \item University of Ottawa (uOttawa)
    \begin{itemize}
    \item Chemical Engineering: 	\url{https://www.uottawa.ca/faculty-engineering/undergraduate-studies/programs/chemical-engineering}
\item Civil Engineering: 	\url{https://www.uottawa.ca/faculty-engineering/undergraduate-studies/programs/civil-engineering}
\item Computer Engineering: 	\url{https://www.uottawa.ca/faculty-engineering/undergraduate-studies/programs/computer-engineering}
\item Electrical Engineering: 	\url{https://www.uottawa.ca/faculty-engineering/undergraduate-studies/programs/electrical-engineering}
\item Mechanical Engineering: 	\url{https://www.uottawa.ca/faculty-engineering/undergraduate-studies/programs/mechanical-engineering}
    \end{itemize}
    
    \item University of Regina (UofR)
    \begin{itemize}
    \item Electronic Systems Engineering:	\url{https://www.uregina.ca/academics/programs/engineering/electronic-systems-engineering.html}
    \item Environmental Systems Engineering:	\url{https://www.uregina.ca/academics/programs/engineering/environmental-systems-engineering.html}
    \end{itemize}
    
    \item University of Saskatchewan (USask)
    \begin{itemize}
    \item Chemical Engineering: 	\url{https://admissions.usask.ca/chemical-engineering.php}
\item Civil Engineering: 	\url{https://admissions.usask.ca/civil-engineering.php}
\item Computer Engineering: 	\url{https://admissions.usask.ca/computer-engineering.php}
\item Electrical Engineering: 	\url{https://admissions.usask.ca/electrical-engineering.php}
\item Environmental Engineering:	\url{https://admissions.usask.ca/environmental-engineering.php}
\item Mechanical Engineering: 	\url{https://admissions.usask.ca/mechanical-engineering.php}
    \end{itemize}
    
    \item University of Toronto (UofT)
    \begin{itemize}
    \item Chemical Engineering: 	\url{https://discover.engineering.utoronto.ca/programs/engineering-programs/chemical-engineering/}
\item Civil Engineering: 	\url{https://discover.engineering.utoronto.ca/programs/engineering-programs/civil-engineering/}
\item Computer Engineering: 	\url{https://discover.engineering.utoronto.ca/programs/engineering-programs/electrical-computer-engineering/}
\item Electrical Engineering: 	\url{https://discover.engineering.utoronto.ca/programs/engineering-programs/electrical-computer-engineering/}
\item Mechanical Engineering: 	\url{https://discover.engineering.utoronto.ca/programs/engineering-programs/mechanical-engineering/}
    \end{itemize}
    
    \item University of Victoria (UVic)
    \begin{itemize}
    \item Civil Engineering: 	\url{https://www.uvic.ca/ecs/civil/undergrad-students/index.php}
\item Computer Engineering: 	\url{https://www.uvic.ca/ecs/ece/future/undergraduate/computer/index.php}
\item Electrical Engineering: 	\url{https://www.uvic.ca/ecs/ece/future/undergraduate/electrical/index.php}
\item Mechanical Engineering: 	\url{https://www.uvic.ca/ecs/mechanical/programs/beng-major/index.php}
    \end{itemize}
    
    \item University of Waterloo (UWaterloo)    \begin{itemize}
    \item Chemical Engineering: 	\url{https://uwaterloo.ca/future-students/programs/chemical-engineering}
\item Civil Engineering: 	\url{https://uwaterloo.ca/future-students/programs/civil-engineering}
\item Computer Engineering: 	\url{https://uwaterloo.ca/future-students/programs/computer-engineering}
\item Electrical Engineering: 	\url{https://uwaterloo.ca/future-students/programs/electrical-engineering}
Environmental	\url{https://uwaterloo.ca/future-students/programs/environmental-engineering}
\item Mechanical Engineering: 	\url{https://uwaterloo.ca/future-students/programs/mechanical-engineering}
    \end{itemize}
    
    \item University of Western Ontario (Western)
    \begin{itemize}
    \item Chemical Engineering: 	\url{https://www.eng.uwo.ca/img/undergraduate/FINAL\_Chemical\_2018.pdf}
\item Civil Engineering: \url{https://www.eng.uwo.ca/img/undergraduate/FINAL\_Civil\_2018.pdf}
\item Computer Engineering: 	\url{https://www.eng.uwo.ca/img/undergraduate/FINAL\_computer\_2018.pdf}
\item Electrical Engineering: 	\url{https://www.eng.uwo.ca/img/undergraduate/FINAL\_electrical\_2018.pdf}
\item Mechanical Engineering: 	\url{https://www.eng.uwo.ca/img/undergraduate/FINAL\_mechanical\_2018.pdf}
    \end{itemize}
    
    \item University of Windsor (UWindsor)
    \begin{itemize}
    \item Civil Engineering: 	\url{https://www.future.uwindsor.ca/program/civil-engineering/}
\item Electrical Engineering: 	\url{https://www.future.uwindsor.ca/program/electrical-and-computer-engineering/}
Environmental	\url{https://www.future.uwindsor.ca/program/environmental-engineering/}
\item Mechanical Engineering: 	\url{https://www.future.uwindsor.ca/program/mechanical-engineering/}
    \end{itemize}
    
    \item York University (YorkU)
    \begin{itemize}
    \item Civil Engineering: 	\url{https://futurestudents.yorku.ca/program/civil-engineering}
\item Electrical Engineering: 	\url{https://futurestudents.yorku.ca/program/electrical-engineering}
\item Computer Engineering: 	\url{https://futurestudents.yorku.ca/program/computer-engineering}
\item Mechanical Engineering: 	\url{https://futurestudents.yorku.ca/program/mechanical-engineering}
    \end{itemize}
    
\end{itemize}

\subsection{Accreditation Boards and Criteria}
\label{app:appendix_signatories}
Our discussion of accreditation criteria focuses on full signatories of the International Engineering Alliance's Washington Accord. Signatories are listed below, along with the acronyms used to identify them throughout this work, and URLs at which we accessed graduate attributes (October 29th, 2026).

\begin{itemize}
    \item Accreditation Board for Engineering and Technology (ABET): 
    \url{https://www.abet.org/wp-content/uploads/2024/11/2025-2026\_EAC\_Criteria.pdf} 
    \item Accreditation Board for Engineering Education of Korea (ABEEK):

    \url{https://www.abeek.or.kr/appraisal/eac}
    \item Association for Engineering Education of Russia (AEER): 
    
    \url{https://aeer.ru/en/accreditation/criteria/}
    \item Association for Evaluation and Accreditation of Engineering Programs (MÜDEK):

    \url{https://www.mudek.org.tr/doc/tr/MUDEK-Degerlendirme\_Olcutleri\_(3.1-05.12.2025).pdf}
    \item Board of Engineers Malaysia (BEM): 
    
    \url{https://bem.org.my/uploads/ISO/Accreditation/67701429b28903cfe8f6126a149066b2.pdf}
    \item China Association for Science and Technology (CAST): \url{https:www.ceeaa.org.cn/upload/files/20251212/613946c27542079f/2022111923195859570.pdf}
    \item Colegio Federado de Ingenieros y de Arquitectos de Costa Rica (CFIA):
    
    \url{https://aapia.cfia.or.cr/wp-content/uploads/2026/08/manual-of-criteria-and-procedures-for-the-accreditation-of-engineering-programs.pdf}
    \item Consejo de Acreditación de la Enseñanza de la Ingeniería (CACEI):
    
    \url{https://cacei.com.mx/wp-content/uploads/2026/02/dac\_01\_doc\_politicas\_procedimientos\_2.pdf}
    \item Engineering Council South Africa (ECSA):

    \url{https://www.ecsa.co.za/download/502/accreditation-policy-and-templates/1007405/e-03-crip-criteria-for-accreditation-of-engineering-programmes}
    \item Engineering Council United Kingdom (ECUK):

    \url{https://www.engc.org.uk/media/isdl1tem/the-accreditation-of-higher-education-programmes-ahep-v44.pdf}
    \item Engineering New Zealand (EngNZ):

    \url{https://d2rjvl4n5h2b61.cloudfront.net/media/documents/ACC\_02\_-\_Accreditation\_Criteria\_V4.3\_1.pdf}
    \item Engineers Australia (EA): 
    
    \url{https://www.engineersaustralia.org.au/sites/default/files/2026-08/national-competency-standard-engineering-2026.pdf}
    \item Engineers Canada (EC):

    \url{https://engineerscanada.ca/sites/default/files/2025-12/2025-Accreditation\_Criteria\_Procedures-EN.pdf}
    \item Engineers Ireland (EI):

    \url{https://www.engineersireland.ie/listings/resource/519}
    \item Indonesian Accreditation Board for Engineering Education (IABEE):

    \url{https://iabee.or.id/download/WWtnNVpsaHBiR3cyWlRZdFVXSXhaVE55V1d0UllUWnpTUzlDUVZKamQyTnZjblpsUkdSRmFWTjBRV0UzZEROcGRtOU5ibEpOTWpWcVRVZFJlbWxEY1hKRmFXbDBRMlU1Y2tocFNXRnNUVEkxYWsxSFVYcEhSVmRPVlVjelRUSTFhazFIMk5DWGtVWHB3VmtObGRYSnpjMkpwVjI5emJsVk5NalZxVFVkUmVqSXllakJPTW13elRXRjNMV2t5U1RWWE1FRXpPVEJ0T0hFd1pqbEZNazgzUlRCVE0xTXhXVGRRTkdJeFpqUlZNREl3VHpoM01rTXpRVGN4TlhJdVduQjVaR2xtU2pNZDNjMG1icjAwMTMy}
    \item Institute of Engineering Education Taiwan (IEET):

    \url{https://www.ieet.org.tw/FileSystem/htmlEditor/c012c1a5-9430-44e9-b252-da6de4923748/B1-1\%20Accreditation\%20Criteria\%20 for\%20Accrediting\%20Engineering\%20Programs.pdf}
    \item  Institution of Engineers Singapore (IES):

    \url{https://www.ies.org.sg/wp-content/uploads/EAB-Accreditation-Manual-Visits-after-July-2024-Final.pdf}
    \item Institution of Engineers Sri Lanka (IESL):

    \url{https://app.box.com/s/a9pk6dhdbejutvl84x279oqiveags2iq}
    \item Instituto de Calidad y Acreditacion de Programas de Computacion, Ingenieria y Tecnologia (ICACIT):

    \url{https://webicacit.com/archivos/2025\_ICACIT\_CriteriosCAI\_eng.pdf}
    \item Japan Accreditation Board for Engineering Education (JABEE): 
    
    \url{https://jabee.org/doc/Category-Discipline-Specific\_Criteria\_2019-r2.pdf}
    \item National Board of Accreditation (India) (NBA):

    \url{https://www.nbaind.org/files/General\%20Manual\%202026\%20Edition\_Format\_20260521170009.pdf }
    \item Pakistan Engineering Council (PEC):

    \url{https://www.pec.org.pk/wp-content/uploads/2021/04/Manual-of-Accreditation-Third-Edition-2019-Amended-Ver.-of-Manual-2014.pdf}
    \item Philippine Technological Council (PTC):

    \url{https://ptc.org.ph/wp-content/uploads/2019/11/CASEE-Manual-V.-2-Accreditation-Criteria-.pdf}
    \item  The Hong Kong Institution of Engineers (HKIE): 

    \url{https://hkie.org.hk/wp-content/uploads/hkie/3c1cd3f952add63df75b174d2680694b/Accr-Handbook-Engg-Degrees-Oct-2024-version-2026.04-v.1.pdf?x38449\&ver=1782106976}
    \item The Institution of Engineers Bangladesh (IEB): 
    
    \url{https://www.baetebangladesh.org/acc-man-02-v2-2-f.html}
    
\end{itemize}

%TC:endignore
\end{document}